\documentclass[a4paper,twocolumn,11pt]{quantumarticle}
\pdfoutput=1

\usepackage{bm}
\usepackage[retainorgcmds]{IEEEtrantools}
\usepackage{graphicx}
\usepackage{mathrsfs}
\usepackage{amsmath}
\usepackage{amssymb}
\usepackage{color}
\usepackage{amsfonts}
\usepackage{times,txfonts}
\usepackage{nicefrac}
\usepackage[colorlinks=true,linkcolor=blue,urlcolor=blue,citecolor=blue,pdfusetitle]{hyperref}
\usepackage{physics}
\usepackage{yquant}
\usepackage{tikz}
\usepackage{subcaption}

\usepackage{booktabs}
\usepackage{tabularx}
\usepackage{multirow}
\usepackage{longtable}
\usepackage{siunitx}
\usepackage{pgfplots}
\usepgfplotslibrary{statistics,groupplots}
\pgfplotsset{compat=1.18}

\definecolor{myGreen}{HTML}{32965D}
\definecolor{myPurple}{HTML}{A546BD}
\definecolor{myRed}{HTML}{B74244}
\definecolor{myBlue}{HTML}{459DD9}

\usepackage[numbers,sort&compress]{natbib}

\usetikzlibrary{shapes.geometric, arrows.meta, positioning}

\usepackage{algpseudocode}
\usepackage{algorithm}
\usetikzlibrary{positioning}
\usepackage{float}

\yquantset{register/default name=$\ket{\reg_{\idx}}$}
\usepackage{todonotes}
\usepackage{marginnote}

\usepackage{layouts}

\begin{document}

\title{Quantum control of solid state qubits via Reinforcement Learning}

\author{Dennis Herb}
\affiliation{Institute for Complex Quantum Systems, Ulm University, Ulm, Germany}
\author{Mirko Rossini}
\affiliation{Institute for Complex Quantum Systems, Ulm University, Ulm, Germany}
\author{Sabine Wölk}
\affiliation{Deutsches Zentrum für Luft- und Raumfahrt e.V. (DLR), Institut
für Quantentechnologien, Ulm, Germany}

\author{Joachim Ankerhold}
\affiliation{Institute for Complex Quantum Systems, Ulm University, Ulm, Germany}
\affiliation{Center for Integrated Quantum Science and Technology (IQST) Ulm-Stuttgart, Germany}
\author{Alessio Belenchia\textsuperscript{*}}
\affiliation{Deutsches Zentrum für Luft- und Raumfahrt e.V. (DLR), Institut
für Quantentechnologien, Ulm, Germany}

\author{Dominik Maile\textsuperscript{*}}
\affiliation{Institute for Complex Quantum Systems, Ulm University, Ulm, Germany}

\begingroup
\renewcommand\thefootnote{\fnsymbol{footnote}}
\footnotetext[1]{These authors contributed equally to this work}
\endgroup

\maketitle

\begin{abstract}
The use of reinforcement learning (RL) in the design of quantum control strategies for pulse level control of NISQ era quantum devices is attracting growing interest. Model-free RL, not requiring preexisting knowledge of the physical system, acts as a black-box optimization tool well suited for noisy devices.
In this work, we provide a simulation-based analysis of the performance of model-free RL for state preparation and qubit gate design in a challenging setup: indirect control of NV nuclear spins in a spin-bath. In particular, the spin-bath is modeled via the generalized cluster correlation expansion accounting for magnetic impurities and intra-bath interactions.
We analyze the performance of the RL agents by performing simulations with minimally observable environments and noisy, sparse rewards as introduced in (Phys. Rev. X 12, 011059), and compare them to other state-of-the-art black-box optimization tools.
\end{abstract}

\setcounter{tocdepth}{2}
\tableofcontents

\section{Introduction}\label{sec:intro}

Optimally controlling quantum computational devices in order to steer their evolution with high fidelity, despite the inherent noise of open quantum systems, is a crucial task to empower quantum computing and technologies. Indeed, high-fidelity state preparation and gate synthesis, together with device characterization, is a prerequisite for making use of current Noisy Intermediate-Scale Quantum (NISQ) devices, empowering quantum sensing, and aiming at fault tolerant machines in the future.

In this context, the rapid advances of machine learning (ML) methods have stimulated the physics community to employ ML methods for the characterization and control of complex quantum systems~\cite{PhysRevX.12.011059,Giannelli2022,gebhart2023learning,krenn2023artificial,Mukherjee2024,alexeev2025artificial,bukov2026reinforcement,kiermeyer2026uncovering}. These methods, together with other classical optimization methods, offer a manifold of tools for ``taming'' quantum systems~\cite{duncan2025taming}.
Among these, reinforcement learning (RL) has emerged as a valuable, competitive and, at times, best performing optimization strategy for state preparation, gate synthesis, error correction, and circuit optimization (see~\cite{bukov2026reinforcement} and references therein).

In typical RL algorithms, the goal is to train an agent to interact with a system, usually referred to as \textit{environment}, with the aim of learning the optimal interaction strategy to drive the environment toward a desired target state.
At each step, the agent observes the environment and selects a permitted action. The action changes the state of the environment and the agent receives a reward based on the quality of the chosen action with respect to the task to accomplish. The process proceeds this way for several steps until a termination condition is met and the training episode ends. The environment is then reset and the process begins anew with the agent exploiting the experience accumulated in previous episodes. In this way the agent learns an optimal strategy, called policy, to maximize the cumulative expected reward thus accomplishing its task(s).

Several different algorithms fall under the umbrella of RL. Among these, there are algorithms particularly suitable for learning in stochastic, noisy environments. This aligns perfectly with the problem of controlling quantum devices, where noise is the limiting factor for precise control.
In order to keep a precise terminology, we will henceforth call the degrees of freedom producing noise for our system/register the \textit{bath}. The RL agent's \textit{environment} is then only partially accessible via the system degrees of freedom, but naturally contains the \textit{system} and the \textit{bath}.

In the context of quantum optimal control, model-free reinforcement learning can be seen as a  black-box optimization method that does not require the characterization of the quantum system nor of its bath.
 The agent acquires information on the full  environment while learning and can be further deployed to control related systems or for slightly different tasks based on its generalization capabilities.
 This represents a great advantage with respect to model-dependent optimization methods whose optimal solutions are highly specialized and can be fragile to un-modeled characteristics of a real quantum system.
However, it is important to note that in many applications of RL to quantum control problems it is assumed that the agent has access to the quantum state (or even the evolution operator) at each step.
Furthermore, the agent also routinely receives rewards that are connected with state or process fidelities which, in general, can only be estimated. While this is not an issue while performing simulations of small systems, the overhead required in real experiments is prohibitively high. Thankfully, strategies to handle minimally observable environments have been considered in the literature~\cite{PhysRevX.12.011059,PhysRevA.105.062443,duncan2025taming,nie2026experiment} and applied to successfully control real quantum computational devices~\cite{sivak2023real,PhysRevX.13.031035}.

In order to theoretically quantify the performance of RL based control in specific open quantum system settings, one has to rely on simulation techniques that can treat system, bath, and driving fields on equal footing. For realistic NISQ devices, the inherent noise typically originates from quantum mechanical interactions which can entangle the system and the noise source. Furthermore, often the frequency spectrum of the bath cannot be assumed to be flat, yielding non-Markovian feedback effects. Such a quantum bath is usually more complex to model and characterize than a Markovian one and its noise more challenging to mitigate while being an important factor in current devices and for the road ahead toward fault tolerance~\cite{PhysRevLett.96.050504,PhysRevLett.110.010502,2024QS&T....9c5027L,white2022characterization,PhysRevX.15.021047}.
Dealing with these complex settings, the Born-Markov approximation is generally not applicable and sophisticated techniques based on the microscopic origin of relaxation and decoherence have to be used for the simulations. This becomes even more apparent in the presence of driving, where the applicability of Lindblad master equations becomes even more limited.

Despite its complexity, the inherent  structure of a bath can also offer possibilities for tailored high fidelity solutions for the control problem at hand. However, without the precise knowledge of the bath such control is generally hard to find. The question whether RL agents are able to navigate within the underlying complex optimization landscape to find novel solutions is therefore of major interest.
Altogether, this makes the understanding of the performance of RL agents in such a complex situation a key insight for future quantum technologies.

\begin{center}
\begin{figure*}[t]
\centering
\includegraphics[width=1.\linewidth]{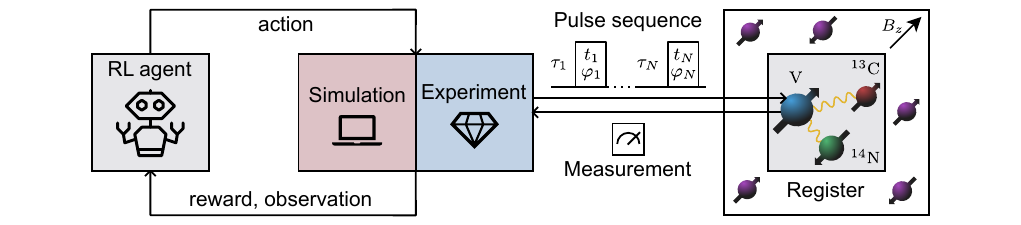}
\caption{\label{fig:overview}
Overview of our RL setup. A reinforcement learning agent indirectly controls one or more nuclear spins via mw pulses on the NV electronic spin while under the influence of a spin-bath. Upon measuring the electronic spin at the end of a train of pulses the agent obtains a reward. The latter aids the agent in learning how to perform pulse sequences aimed at the optimal implementation of quantum gates or state preparation while accounting for the open quantum system dynamics.}
\end{figure*}
\end{center}
In this work, we employ RL to design pulse sequences for controlling the dynamics of nuclear spins in a nitrogen-vacancy (NV) center while accounting for its microscopic spin-bath (see Fig.~\ref{fig:overview}). NV centers in diamond are, together with other solid state defects, at the forefront of quantum sensing applications; they are promising systems for solid state quantum memories and networks, and have been deployed as quantum processors~\cite{rondin2014magnetometry,PhysRevX.9.031045,pompili2021realization,stolk2024metropolitan,childress2013diamond,abobeih2022fault,chang2025hybrid} (and references therein).
Quantum registers in diamond are realized by coupling an NV center electronic spin to surrounding, addressable nuclear spins. While the central electronic spin offers fast control and high-fidelity readout via optical fields, the nuclear spins sport high coherence times suitable for quantum memories and can be indirectly controlled via the electronic spin -- or directly via radio frequency (rf) fields. At the same time, the rest of the unaddressed nuclear spins form a spin-bath that makes up the dominant magnetic noise source giving rise to decoherence and gate errors. Within such registers, quantum states can be prepared and manipulated through quantum gate operations implemented by carefully designed pulse sequences.

We use model-free RL to design robust pulse sequences for gate synthesis assuming minimal observability of the agent's environment and employing resource-realistic reward schemes for the agents.
In order to model the NV center registers and their surrounding spin-bath microscopically we exploit the generalized cluster correlation expansion (gCCE) technique~\cite{PhysRevB.110.035407}.
This method allows us to go beyond a simple mean field treatment of the spin-bath and, at the same time, retain the bath's microscopic structure and memory which are lost in a Lindblad master equation approach. This is crucial to model central-spin systems where the bath is a sparse set of interacting nuclear spins.
Thanks to the gCCE we can efficiently simulate the fidelity of quantum processes, used for rewarding our agents, while accounting for magnetic impurities and spin-spin interactions in the bath. In this way, we train our agents on realistic models of NV registers.
This is combined with sim-to-real transfer techniques from the RL literature~\cite{zhao2020sim} for generating robust solutions. These techniques were originally motivated by robotics applications of RL where collecting real-world data is prohibitively expensive/slow. With these, RL solutions can be either immediately transferred to physical devices or gain a head start in learning that allows for fine-tuning on physical devices in realistic timescales.

As a central result, we find that the RL agents are able to take the specific structure of the spin-bath into account. This yields tailored pulse sequences with higher fidelities compared to benchmark solutions we obtain with other state-of-the-art black-box optimization methods or the same agents trained without the spin-bath.

The work is organized as follows. In section~\ref{sec:NV}, we define the full Hamiltonian describing the NV center registers and their spin-bath. Here we also introduce the gCCE and define the indirect control problem that we investigate via RL. In section~\ref{sec:RLtheory} we introduce the RL algorithm used with particular attention to its sparse reward strategy. Section~\ref{sec:results} summarizes our main results and section~\ref{sec:bench} compares with state-of-the-art black-box optimization methods other than model-free RL. We conclude in section~\ref{sec:conclusions} with a discussion of our findings and outlook for future investigations.

\section{NV quantum registers}\label{sec:NV}
We consider a multiqubit register consisting of the NV center -- electron spin and the nitrogen ${}^{14}$N nuclear spin -- together with a set of ${}^{13}$C nuclear spins coupled to the electronic one.
Generally, the spin-bath contains the spins that are not included in the register, among which are e.g., weakly coupled ${}^{13}$C nuclear spins, P$1$ centers, and unbound electron spins on the diamond surface. Here we solely focus on the $^{13}$C nuclear spin-bath which is unavoidable for $^{13}$C based registers and dominant for the abundances considered here~\cite{PhysRevB.85.115303}.

Following the notation in~\cite{PhysRevB.110.035407}, we introduce the relevant spin operators $\mathbf{S}$ for the NV electronic \mbox{spin-1}, $\mathbf{I}_N$ for the nitrogen nuclear spin-1, $\mathbf{I}_C$ for the ${}^{13}$C nuclear spin-1/2 in the register, and $\boldsymbol{\mathcal{E}}_C$ for the ${}^{13}$C nuclear spin-1/2 in the bath.
Except for the NV center electron spin and the nitrogen spin, spin matrices are defined as the Pauli matrices (including the prefactor 1/2). The spin matrices for the NV center electron spin are defined as spin-1 matrices restricted to the subspace of $m_s=0$ and $m_s=-1$
\begin{align*}
S_x &= \frac{1}{\sqrt{2}}
\begin{pmatrix}
0 & 1 \\
1 & 0
\end{pmatrix}, \quad
S_y = \frac{1}{\sqrt{2}}
\begin{pmatrix}
0 & -\mathrm{i} \\
\mathrm{i} & 0
\end{pmatrix}\\
S_z &=
\begin{pmatrix}
0 & 0 \\
0 & -1
\end{pmatrix}.
\end{align*}
The full Hamiltonian of the system is given by
\begin{align}
    H &= H_{R} + H_{B} + H_{I},
\end{align}
where $H_R$ is the register's Hamiltonian, $H_B$ is the bath Hamiltonian, and the register-bath interaction is $H_{I}$.
The interactions between the different spin pairs are of dipolar nature and have the form
\begin{align}
     \mathbf{S}_1 \mathrm{\mathbf{M}} \mathbf{S}_2 = \frac{\mu_{0}}{4 \pi} \frac{\hbar\gamma_{1} \gamma_{2} }{r^{3}} \left[ \mathbf{S}_1 \cdot \mathbf{S}_2 - 3(\mathbf{S}_1 \cdot \mathbf{n})(\mathbf{S}_2 \cdot \mathbf{n}) \right],
\end{align}
in angular frequency units.
$\mathrm{\mathbf{M}}$ is the hyperfine coupling tensor and encodes the strength of the dipolar magnetic interaction between spins while $\mathbf{n}=\mathbf{r}/|\mathbf{r}|$ with the vector $\mathbf{r}$ connecting the spin positions. The gyromagnetic factors of the electron spin, the nuclear carbon spin, and nuclear nitrogen spin are given by $\gamma_e = -176~\mathrm{GHz/T}$, $\gamma_C = 67.3~\mathrm{MHz/T}$, $\gamma_N = 19.3~\mathrm{MHz/T}$ respectively.

The register Hamiltonian $H_R=H_{NV}+H_{N}+H_{C,R}+H_{I,R}$ contains contributions from the NV center electron spin $H_{NV}$, nitrogen ${}^{14}$N nuclear spin $H_{N}$, and the ${}^{13}$C nuclear spins $H_{C,R}$, together with their interactions $H_{I,R}$,
\begin{align}
&H_{\mathrm{NV}}/\hbar
&&= 2\pi D_{\mathrm{gs}} S_z^2
 + \gamma_e \mathbf{B}\cdot \mathbf{S}, \nonumber\\
&H_N/\hbar
&&= 2\pi P_{\mathrm{gs}} I_{N,z}^2
 + \gamma_N \mathbf{B}\cdot \mathbf{I}_N, \nonumber\\
&H_{C,R}/\hbar
&&= \sum_k \gamma_C \mathbf{B}\cdot \mathbf{I}_C^{(k)}, \nonumber\\
&H_{I,R}/\hbar
&&= \mathbf{S}\mathbf{M}_{NV-N}\mathbf{I}_N
 + \sum_k
 \mathbf{S}\mathbf{M}^{(k)}_{NV-C}
 \mathbf{I}^{(k)}_C \nonumber\\
&&&\quad
 + \sum_k
 \mathbf{I}_N \mathbf{M}^{(k)}_{N-C}
 \mathbf{I}^{(k)}_C \nonumber\\
&&&\quad
 + \sum_{k<l}
 \mathbf{I}^{(k)}_C
 \mathbf{M}^{(k,l)}_{C-C}
 \mathbf{I}^{(l)}_C .
\end{align}
Here $D_{\rm gs} = 2.87~\mathrm{GHz}$ is the zero field splitting of the triplet electronic state, $\mathbf{B}$
is an external magnetic field, and $P_{\rm gs} = -5.08~\mathrm{MHz}$ is the quadrupole splitting of the nitrogen spin.
The bath Hamiltonian $H_B=H_{C,B}+H_{I,B}$ contains the ${}^{13}$C bath spins $H_{C,B}$ with their interactions $H_{I, B}$:
\begin{align}
    {H_{C, B}}/\hbar &= \sum_j \gamma_C \mathbf{B}\cdot \boldsymbol{\mathcal{E}}_C^{(j)} \\
    H_{I, B}/\hbar &= \sum_{i<j} \boldsymbol{\mathcal{E}}^{(i)}_C\mathbf{M}^{(i,j)}_{C-C} \boldsymbol{\mathcal{E}}^{(j)}_C.
\end{align}
Finally, the interaction Hamiltonian between bath and register reads
\begin{align}
H_I/\hbar
&= \sum_j
\mathbf{S}\mathbf{M}^{(j)}_{NV-C}
\boldsymbol{\mathcal{E}}^{(j)}_C
\nonumber\\
&+ \sum_j
\mathbf{I}_N \mathbf{M}^{(j)}_{N-C}
\boldsymbol{\mathcal{E}}^{(j)}_C
\nonumber\\
&+ \sum_{k,j}
\mathbf{I}^{(k)}_C
\mathbf{M}^{(k,j)}_{C-C}
\boldsymbol{\mathcal{E}}^{(j)}_C .
\end{align}
Notice that, given the positions -- or, alternatively, the dipolar coupling strengths which are usually measured in experiments -- and spin types the full Hamiltonian can be constructed.

In the following, we restrict to the case in which the nitrogen spin stays fully polarized in its $m_N = -1$ state
and it is thus only  renormalizing the magnetic field for the electron spin.
Furthermore, since the zero field splitting $D_{gs}$ of the electron spin is several orders of magnitude larger than the Zeeman splitting of the $^{13}$C nuclear spins and the dipolar coupling to them, the excitation of the electron spin is not affected by the bath. Hence, we can neglect the terms containing $S_x$ from the dipolar interaction. Using this -- and the fact we are only driving the electron spin on resonance -- we transform the Hamiltonian into the rotating frame of the drive and find for the $^{13}$C nuclear spin register the system Hamiltonian~\cite{PhysRevLett.124.220501, PhysRevB.110.035407}

\begin{align}
\bar{H}_R
=& \ket{0}\bra{0} \otimes H_{C,R}^{(0)}
 + \ket{-1}\bra{-1} \otimes H_{C,R}^{(-1)},
\end{align}
where $\ket{0}$ and $\ket{-1}$ are the eigenstates for the $m_s=0$ and $m_s=-1$ eigenvalues, respectively. The Hamiltonians for the nuclear spins in the corresponding subspace of the NV-center spin read
\begin{align}
H_{C,R}^{(0)}/\hbar
=& \sum_k \gamma_C B_z I_{C,z}^{(k)}    + \sum_{k<l}
 \mathbf{I}^{(k)}_C
 \mathbf{M}^{(k,l)}_{C-C}
 \mathbf{I}^{(l)}_C , \nonumber\\
H_{C,R}^{(-1)}/\hbar
=& \sum_k
\left(\gamma_C B_z + M_{zz}^{(k)}\right) I_{C,z}^{(k)}\\
  +& \sum_k M_{zx}^{(k)} I_{C,x}^{(k)} + \sum_k M_{zy}^{(k)} I_{C,y}^{(k)}\nonumber\\
   +& \sum_{k<l}
 \mathbf{I}^{(k)}_C
 \mathbf{M}^{(k,l)}_{C-C}
 \mathbf{I}^{(l)}_C .
\nonumber
\end{align}
Here, the quantization axis of the $^{13}$C spins in the $\ket{-1}$ subspace can be tailored by engineering the magnetic field $B_z$, as indicated by the first bracket in $H_{C,R}^{(-1)}$. We utilize this as a control parameter in the following.

\subsection{Indirect control problem}\label{sec:indirect}
There is a vast literature on how to control the dynamics of the spins in an NV center based register. When selective electron spin resonances (ESRs) are not an option -- e.g., when aiming at controlling weakly coupled nuclear spins -- state-of-the-art methods comprise dynamical decoupling (DD) sequences targeting the sole electronic spin~\cite{PhysRevLett.109.137602}, DDrf methods~\cite{PhysRevX.9.031045} that supplement the DD sequence with rf pulses targeting the nuclear spins, and indirect control with optimized few pulse sequences~\cite{PhysRevA.98.042302,PhysRevLett.124.220501,PhysRevApplied.12.064047,PhysRevLett.125.030501}.

Indirect control of the $^{13}$C spins with few short mw pulses and delays has been analyzed in~\cite{PhysRevA.98.042302,PhysRevLett.124.220501,PhysRevApplied.12.064047} as a means for (fast) controlling weakly coupled nuclear spins, without rf pulses acting directly on the nuclear spins, while also avoiding a costly large number of control operations.
There, the NV spin is controlled via microwave (mw) fields described by the Hamiltonian
\begin{align}
    H_{\rm mw}(\phi)/\hbar = \Omega (\cos\phi S_x + \sin\phi S_y),
\end{align}
where the Rabi frequency $\Omega$ defines the rotation angle when multiplied by the time duration of the mw pulse $t$ and the angle $\phi$ defines the rotation axis in the equatorial plane of the NV center's Bloch sphere.

We consider pulse sequences interleaved with free evolution periods described by the operator
\begin{align}\label{eq:parop}
     \mathcal{S}(\boldsymbol{\tau}, \boldsymbol{t}, \boldsymbol{\phi}) = \prod_{i} U(t_i,\phi_i) U(\tau_i),
\end{align}
where $\tau_i$ are free evolution times and the pulse propagator is given by
\begin{equation}
    U(t_i,\phi_i) = \exp{\left[-i t_i(H_{R}+H_{\rm mw}(\phi_i))/\hbar\right]},
\end{equation}
with the pulse phase $\phi_i$ and duration $t_i$.

Furthermore, as detailed in the following, we can additionally require our agents to learn not only the best pulse parameters for the indirect control, but also the magnitude of the external magnetic field $\mathbf{B}=B_z\mathbf{e}_z$. In this way we train agents that are able to inform experimenters about both the best pulse sequence and the best experimental value of the external magnetic field applied to the system for gate design. Finally, when considering the challenging problem of generating entanglement between two nuclear spins, we will include in the evolution instantaneous  $\pi$-pulses in between the free evolution times as solutions of this form presenting high-fidelity have been considered in the literature~\cite{PhysRevB.110.035407}.

\subsection{Spin-bath modeling: Generalized Cluster Correlation Expansion}

Here we consider how the spin-bath is treated in order to simulate the open quantum dynamics of the register. Due to the exponentially growing Hilbert space of the register-bath system, an exact simulation of the spin-bath is only possible for small spin-baths.  Although we consider only five spins in the bath in this work -- distributed within a sphere of radius $r_{\rm max}=1.3~\mathrm{nm}$ corresponding to an average abundance of $\eta\approx 0.3$\% -- we find that the computational time for exact diagonalization of the system+spin-bath is still too long to train the RL agent for a sufficient number of episodes. We thus rely on  methods approximating the effect of the spin-bath on the register which are also scalable.

The generalized cluster correlation expansion (gCCE) incorporates the microscopic structure of the bath and bath interactions to different orders~\cite{PhysRevB.110.035407}.
\begin{figure}[t]
\centering
\includegraphics[width=1\linewidth]{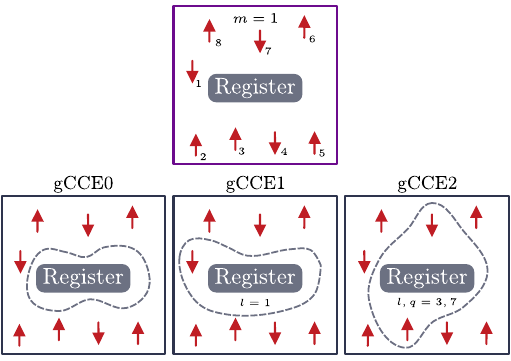}
\caption{\label{fig:gCCE}
Schematic representation of the register and the first gCCE orders. (Top) The register, composed of the electronic spin and selected nuclear spins, is surrounded by impurity spins forming the spin-bath. (Bottom) Illustration of gCCE0, gCCE1, and gCCE2. The encircled cluster contains the register together with the bath spins included at the respective gCCE order, while all remaining bath spins are treated as a mean field~\cite{10.1016/j.aop.2019.168063}.}
\end{figure}
In the gCCE, clusters of the register and selected bath spins are constructed (see Fig.~\ref{fig:gCCE}). The number of ${}^{13}$C {bath nuclear spins whose full interaction with the register is considered}
for each cluster determines the gCCE order. The interaction between the remaining bath spins and the cluster is approximated in a mean-field fashion, i.e., bath spins are no longer treated as spin operators but the expectation values of the z-component $\langle \mathcal{E}_{C,z}^{(l)}\rangle$ are chosen as $\pm 1/2$  according to a classical probability distribution. This implements the completely mixed thermal bath state $\rho_B = \mathbb{I}_B/2^{N}$, where $N$ is the number of spins in the bath.

Since we are interested in the indirect quantum control of the register by pulse sequences interleaved with free evolution times, we consider acting on the state of the system via propagators parametrized as in Eq.~\eqref{eq:parop}.
Then, the $n$-th order gCCE contribution to the density matrix is obtained as follows.
At the zeroth order, for a given bath configuration labeled by the index $m$, we evolve the initial state $\rho(0)$ of the register as $\rho_{m}^{\rm (E0)}(t) =  \mathcal{S}_m \rho(0) \mathcal{S}_m^\dag$. Here the Hamiltonian entering the free evolution propagator -- and pulse propagators when considering finite duration pulses -- contains the Hamiltonian $H^{\rm(E0)}_{m} = H_R + H^{\rm(E0)}_{B,m}+H^{\rm(E0)}_{I,m}$ where the last two terms use the mean field approximation for the bath spins. Moving to the first order contributions (gCCE1), we have the initial state of the register and bath spin $l$ given by $\sigma_{m,l}^{(\mathrm{E1})}(0) = \rho(0) \otimes \rho_{m;l}(0)$ evolving to $\sigma_{m;l}^{(\mathrm{E1})}(t) = \mathcal{S}_{m;l}\, \sigma_{m;l}^{(\mathrm{E1})}(0)\, \mathcal{S}_{m;l}^\dagger$. Here, the Hamiltonian entering the evolution operator contains also the full interaction between the register and bath spin $l$ as well as the bath spin local Hamiltonian. We then trace out the bath spin to recover the state of the register $\rho_{m;l}^{\rm (E1)}(t) = \mathrm{Tr}_{B,l} \left( \sigma_{m;l}^{(\mathrm{E1})}(t) \right)$. An element of the density matrix of the gCCE1 contribution is then obtained by repeating this procedure for all bath spins and using
\begin{align}
   \bra{a} \rho^{\mathrm{(E1)}}_{m} (t)  \ket{b} &=
   \bra{a} \rho_{m}^{\mathrm{(E0)}}(t)\ket{b}\,
   \prod_l  \frac{\bra{a}\rho_{m,l}^{\mathrm{(E1)}}(t)\ket{b}}  {\bra{a}\rho_{m}^{\mathrm{(E0)}}(t)\ket{b}}.
\end{align}
Analogously, at the second order we evolve the state of the register and two bath spins $l,q$ via $\rho_{m,l,q}^{\rm (E2)}(t) = \mathrm{Tr}_{B,l,q} \left(\mathcal{S}_{m;l,q}\, \rho(0)\otimes\rho_{m;l,q}(0)\, \mathcal{S}^{\dagger}_{m;l, q} \right)$ and use
\begin{align}\label{eq:gCCE}
\bra{a} \rho^{\mathrm{(E2)}}_{m}(t)&\ket{b}
= \bra{a} \rho_{m}^{\mathrm{(E1)}}(t)\ket{b}
\nonumber\\
&\quad\times
\prod_{l,q}
\frac{
\bra{a} \rho_{m,l,q}^{\mathrm{(E2)}}(t) \ket{b}
\bra{a} \rho_{m}^{\mathrm{(E0)}}(t) \ket{b}
}{
\bra{a} \rho_{m,l}^{\mathrm{(E1)}}(t) \ket{b}
\bra{a} \rho_{m,q}^{\mathrm{(E1)}}(t) \ket{b}
}.
\end{align}
We proceed in this way until the desired gCCE order (see Appendix C of Ref.~\cite{PhysRevB.110.035407}) and, at the end, we average over $M$ different spin-bath configurations and sum the contributions of the gCCE orders
\begin{align}
    \rho(t) \!=\! \frac{1}{M}\sum_{m=1}^M \left(\rho^{\mathrm{(E0)}}_m(t) \!+\! \rho^{\mathrm{(E1)}}_m(t) \!+\! \rho^{\mathrm{(E2)}}_m(t) \!+\! \dots\!\right).
\end{align}

{In the following, after training an RL agent within a spin-bath approximated by the gCCE, we test the pulse sequences found on the full spin-bath via exact diagonalization. This allows us to verify a posteriori the used order of the cluster expansion. Furthermore, it also highlights the amount of correlation in the spin-bath the RL agent was able to take into account. This becomes especially interesting for long sequences, where the interactions between nuclear spins in the spin-bath become important.}

\subsubsection{Tracing out the NV center}

When interested only in the dynamics of the nuclear spin register we consider solely its reduced state obtained by tracing out the NV
center as $\bar{\rho}(t) = \mathrm{Tr}_{\rm NV}\{\rho(t)\}$.
However, evaluating Eq.~(\ref{eq:gCCE}) in full and tracing out afterward the NV spin is
numerically unstable, since it leads to the propagation of NV degrees of freedom that are
subsequently discarded.
We therefore choose to trace out the NV center at the level of the initial state and propagate the reduced register directly.

Starting from the reduced initial state
$\bar{\rho}(0) = \mathrm{Tr}_{\rm NV}\{\rho(0)\}$, the zeroth-order dynamics is
represented via
a reduced CPTP channel with Kraus operators
$K^\alpha_{m} = \bra{\alpha}\mathcal{S}_{m}\ket{0}$,
\begin{align}
    \bar{\rho}_{m}^{(\mathrm{E0})}(t) =
    \sum_{\alpha=1,2} K^\alpha_{m}\,
    \bar{\rho}(0)\,
    \big(K^{\alpha}_{m}\big)^\dagger ,
\end{align}
where $\ket{0}$ and $\ket{\alpha}$ denote the initial and final NV states,
respectively. Analogously, the first-order contributions start from
$\bar{\sigma}_{m;l}^{(\mathrm{E1})}(0) = \bar{\rho}(0) \otimes \rho_{m;l}(0)$ and
are propagated with the Kraus operators
$K^\alpha_{m;l} = \bra{\alpha}\mathcal{S}_{m;l}\ket{0}$. As in
Eq.~(\ref{eq:gCCE}), the reduced register dynamics $\bar{\rho}(t)$ is then
reconstructed by combining the different gCCE orders.

\section{Reinforcement learning for pulse sequence design}\label{sec:RLtheory}
A common problem of many quantum optimization methods is model bias, i.e., the dependence of the optimized solution on a model of the system under investigation. This is particularly relevant for gradient-based quantum control, where gradient information has to be inferred from the model and, more generally, for classical optimization routines dependent on simulators that cannot capture the full breadth of the real systems' complexity.

While a model-free RL agent does not require any information on the system \textit{a priori}, the way in which agents have been employed for quantum gate design and state preparation has been widely based on simulators which are subject to model bias. Moreover, in most studies about the applicability of model-free RL to optimal control problems, the agent observes at each step the quantum state of the system -- or, even, the propagators of the physical process -- and collects a reward proportional to some fidelity measure or, at least, does so at the end of each episode to collect a final reward. Observing the state, or propagator, and collecting the reward requires tomographic information (see e.g.,~\cite{PRXQuantum.2.040324}) which, in the best case makes the process prohibitively expensive in a realistic scenario and, in the worst case, makes it non-applicable in a real device. Indeed, observing the system disrupts the same evolution one aims at controlling. Thus, to obtain a model-free RL agent that can be, in principle, trained on a real device one needs to account for minimally observable environments and scarcity of detailed rewards.

Fortunately, recent advancements have tackled this issue by showing that RL agents with sparse reward strategies are able to learn even when only presented with single(few)-shot(s) reward estimators from quantum measurements and in the presence of minimally observed environments~\cite{PhysRevX.12.011059,nie2026experiment}. Some of these results have been successfully employed in systems based on superconducting qubits for two-qubit gate design and implementation of the Gottesman-Kitaev-Preskill (GKP) error correction code~\cite{sivak2023real,PhysRevX.13.031035}. In this work we build upon these results and show that agents can learn strategies for indirect control of nuclear spins in the presence of a structured bath with minimal information.

\subsection{RL specifications}
In RL algorithms, the agent selects actions which in turn modify the state of its environment, i.e. the physical system it interacts with. It then acquires rewards depending on the action taken and the subsequent state of the environment with the objective to maximize the sum of expected rewards. The learning proceeds in episodes divided into time steps. At each time step the agent selects an action and (can) receive a reward. Trying to maximize the expected return in an episode, i.e., the predicted total future reward that the agent expects, the agent learns optimal strategies to accomplish its task. The end result is a trained policy that determines the actions the agent needs to take when deployed in the environment for achieving its objective.

In our case the environment of the agent is the NV center system itself together with its spin-bath -- or the simulation of it. The agent acts on the environment via the application of mw pulses to the NV electronic spin to accomplish state preparation or gate design tasks.
Thus, in this work we deal with a stochastic environment and use a continuous state space. Here, the states of the agent's environment can in principle be either the quantum states of the system or the time-propagators.
{However, while for simulated environments passing to the agent the quantum state of the system or its time propagator at each time step in the episode is easily done, doing so would invalidate the claim that the RL method used is \textit{model-free}.} Furthermore, it is clearly unrealistic to expect the agent to have the knowledge of the quantum state -- or even the propagator -- for deployment in realistic setups.
We thus consider a partially observable environment where the agent gets observations with partial information instead of the full state of the environment (cf. Appendix~\ref{app:RL}).
For the tasks of gate and state design via pulse sequences that we consider in this article, the action space is also continuous as the mw pulses and free evolution are  parametrized via their duration and phase.

 Given these specifications, the Proximal Policy Optimization (PPO) algorithm~\cite{schulman2015high} becomes a natural choice for the problem at hand. PPO is an actor-critic method that can be used for problems with continuous action and state/observation spaces and it trains a stochastic policy that gives rise to a probability distribution over the action space. PPO performs particularly well in stochastic environments due to the stability coming from discouraging large policy updates.
 We employ the PPO implementation from Stable-Baselines3 framework~\cite{stable-baselines3} with a custom environment adapted to the physical system under consideration\footnote{We refer the interested reader to the Stable-Baselines3 documentation \cite{stable-baselines3} for the details on the PPO algorithm implementation. Stable-Baselines3 is a \textit{set of reliable implementations of reinforcement learning algorithms in PyTorch} developed and maintained at DLR. It is a user friendly, low-access point for developing RL pipelines. Note that, using a recurrent version of PPO employing LSTM neural network layers is possible via the SB3 Contrib project.}. {Note that, contrary to~\cite{PhysRevX.12.011059} where a recurrent version of PPO is used, we observe effective learning while using the base version of PPO (no LSTM layer is included).} Learning without an LSTM layer is computationally much faster and, for the tasks at hand, does not impact the quality of the solutions considerably.

The main elements of our framework are:

\begin{itemize}
    \item[(I)] \textit{Action space:} normalized $n$-dimensional box $[-1,1]^{\otimes n}$. For most of our examples, at each step the agent chooses three action parameters that represent -- once properly rescaled -- the idle time, the pulse duration and the pulse phase.
    Moreover, we include the option of letting the agent choose the magnitude of the external magnetic field at the start of each episode to show that the agent is also able to learn the optimal magnetic field for indirect control. {It is important to notice that, in the way the normalized actions are rescaled lies the expert-knowledge of the problem at hand that goes into the engineering of the RL agent.} The same expert knowledge is exploited in all the other black-box optimization methods we consider in section~\ref{sec:bench};
    \item[(II)] \textit{Observation space:} since we aim at training hardware friendly agents, we employ a one-hot encoding of the time step in the episode as in~\cite{PhysRevX.12.011059}. This corresponds to an agent that essentially does not get any observation apart from being aware of the elapsed time~\cite{pardo2018time}. While time alone has proven to be sufficient for learning, we also often augment the observation with the previous action taken by the agent;
    \item[(III)] \textit{Episode:} The agent aims at constructing a     parametrized circuit by adding at each step in an episode a layer   composed of an idle time and a mw pulse. Episodes have a finite horizon dictated by the chosen number of pulses for the pulse sequence.
    \item[(IV)] \textit{Reward:} In all cases we consider the reward as sparse, i.e., it is assigned only at the end of an episode.
    The reward schemes that we consider are:
    \begin{itemize}
        \item \textit{Ideal:} considered in most of the literature, it assigns the relevant exact fidelity as the reward at the end of an episode. In a laboratory setting this would require state or process tomography at the end of each episode;
        \item \textit{Single(few)-shot(s) fidelity estimate:} following~\cite{PhysRevX.12.011059}, we can assign as reward at the end of an episode a stochastic quantity corresponding to the single-shot result of a measurement as long as, on average, this quantity gives an estimator of the fidelity for the task considered (cf. Sec.~\ref{sec:rewards}).
    \end{itemize}
\end{itemize}

The hyperparameter specification for our trained agents can be found in Appendix~\ref{app:hyperpa}. Note that we did not embark on an extensive hyperparameter fine-tuning. This means that, in principle, the performance of our agents could be further increased for some of the tasks discussed in this article. We leave this for future works aiming at being deployed on real devices.

\subsection{Sparse reward strategies}\label{sec:rewards}
We discuss now which fidelity estimates we use in more detail. It is crucial to note that, all single-shot rewards that we discuss in the following only require simple Pauli measurements and separable state preparation, an important feature when it comes to RL training on an actual hardware.

\subsubsection{State engineering}
When considering state engineering the figure of merit that we aim at maximizing is the state fidelity.
Consider the case in which we want to prepare a target pure state with density matrix $\rho$. The state fidelity between this target state and a state $\tilde{\rho}$ (which is, in general, a mixed state) is given by
\begin{equation}
    F=\mathrm{Tr}(\rho\tilde{\rho})= \sum_{i=1}^{d^2} \mathrm{Tr}[\rho P_i] \mathrm{Tr}[\tilde{\rho} P_i]/d,
\end{equation}
where $d=2^n$, $n$ is the number of qubits, and $P_i$ are the Pauli operators with $\mathrm{Tr}[P_i P_j]= d \delta_{ij}$.  We show in Appendix~\ref{app:statefid} that this can be estimated via
\begin{equation}
    F\sim \frac{1}{d\, N\, M}\sum_n\sum_m \frac{1}{\mathrm{Pr}(i_n)}\mathrm{Tr}[\rho P_{i_n}]e^{(i_n)}_m,
\end{equation}
where $P_i$ are Pauli operators defining the measurement scheme, $\mathrm{Pr}(i)=\mathrm{Tr}[\rho P_i]^2/d$ is the relevance distribution telling us how relevant a certain Pauli operator is in determining the state fidelity, and the binary eigenvalues $e^{(i_n)}_m$ -- resulting from the measurement of Pauli operators in a separable eigenbasis -- are the information that we extract in a single-shot measurement.
Then the single-shot reward for the RL agent is given by $1/d \mathrm{Tr}[\rho P_{i_n}]e^{(i_n)}_m/{\mathrm{Pr}(i_n)}$ and, on average, it returns a faithful estimator of the state fidelity. In particular, this can be easily implemented in the lab via local Pauli measurements by just adding a layer of local single qubit rotations and measuring in the computational basis.
Note moreover that the last expression allows us to also consider more than a single shot and thus use a better, but still feasible, state fidelity estimator as reward. It should be noted that in~\cite{PhysRevLett.106.230501} a prescription for the number of shots that depends on the measurement is defined. Here we have not followed this scheme since it is not relevant for the single-shot reward. However, when considering more than one shot it could be ideal to look at a variable number of shots for each Pauli operator.

\subsubsection{Gate engineering}
When considering gate engineering the figure of merit that we aim at maximizing is the process fidelity.
Consider for simplicity the case of a target unitary $U$ and an imperfect unitary $V$ that we wish to optimize -- the general case where $V$ is replaced by a CPTP channel is straightforward.
The process fidelity is
\begin{equation}
F=|\mathrm{Tr}(U^\dag V)|^2/d^2.
\end{equation}
This expression can be readily used as the sparse reward for our RL agent.
However, when a laboratory feasible estimator is needed we can instead follow different routes.

\paragraph{Full direct fidelity estimation (DFE):}

The process fidelity between a unitary gate and a quantum channel is expressed as $F=\mathrm{Tr}[\mathcal{U}^\dag \Lambda]/d^2$ where here $\mathcal{U}$ and $\Lambda$ represent the superoperator matrices of the target gate and the experimental channel respectively~\cite{PhysRevLett.106.230501}.

This can be rewritten as
\begin{equation}\label{DFE1}
    F=\frac{1}{d^2} \sum_{k,k'=1}^{d^2}\chi_\mathcal{U}(k,k')\chi_\Lambda(k,k'),
\end{equation}
where $\chi_\mathcal{U}(k,k')= \mathrm{Tr}[P_k \mathcal{U}(P_{k'})]/d$, with $\{P_k\}$ the Pauli operators basis, and analogously for $\Lambda$. As we detail in Appendix~\ref{app:dfe}, this fidelity can be estimated via
\begin{equation}\label{DFE41}
    F\sim \frac{1}{N\,M\, L}\sum_{i=1}^N\sum_{m=1}^M\sum_{m'=1}^L \left[\frac{e_{m}^{k'_i}e_{m'}^{k_i}}{\chi_\mathcal{U}(k_i,k_i')}\right].
\end{equation}
In this expression, $e_{m'}^{k_i}$ is the binary outcome of a local measurement over the separable state eigenbasis of Pauli operator $P_{k_i}$, while $e_{m}^{k'_i}$ is related to the initial separable input state in the current experiment.
We then see that the single-shot reward takes the form $[{e_{m}^{k'_i}e_{m'}^{k_i}}/{\chi_\mathcal{U}(k_i,k_i')}]$.
Note that in~\cite{PhysRevLett.106.230501} a prescription for the number of shots that depends on the measurement and input states is defined. {Again, as for the case of state fidelity, we do not follow this prescription here since it is not relevant for the single-shot reward.}
\\
\\
\indent Since the process fidelity is notoriously expensive to estimate, many alternatives have been proposed in the literature. We thus briefly discuss the single-shot formulation of the widely used lower bound to the process fidelity proposed by Hofmann in~\cite{PhysRevLett.94.160504} while in Appendix~\ref{app:0fid} we consider also the 0-fidelity~\cite{PhysRevResearch.3.033031} which has been already successfully used for quantum optimization purposes.

\paragraph{Hofmann fidelity bounds:}
A lower bound to the process fidelity is given in~\cite{PhysRevLett.94.160504} in terms of two classical fidelities. The bound reads
\begin{equation}
    F\geq F_1+F_2-1
\end{equation}
where $F_{i}=\sum_{n=1}^{d} F_{i,n}/d$ with $i\in \{1,2\}$ and
\begin{equation}
    F_{i,n} = \mathrm{Tr}(\mathcal{U} (\ket{n}_i\bra{n})\,\,\Lambda(\ket{n}_i\bra{n})).
\end{equation}
In the last expression the states $\ket{n}_i$ are chosen as the computational basis for $F_1$ and the $x$-rotated basis for $F_2$.
When only local measurements of Pauli operators are available we can expand the lower bound expression to arrive at (cf. Appendix~\ref{app:hoffman})
\begin{align}
    F_1+F_2  &\sim \frac{2}{d\, N\, M\, L}\\&\times \nonumber \sum_{n,m,l}\frac{1}{\mathrm{Pr}(j_{n,m})}\mathrm{Tr}[\rho_n P_{j_{n,m}}]e^{j_{n,m}}_l.
\end{align}
Here $\mathrm{Pr}(j_{n,m})=\mathrm{Tr}[\rho_n P_{j_{n,m}}]^2/d$ is the relevance distribution telling us how relevant a certain Pauli operator is in determining the state fidelity for a particular $n$.
This fidelity estimator is built from the results of local Pauli measurements in an experiment where the initial state is drawn uniformly at random in the set of states considered. Estimating such lower bound is, in general, more convenient in terms of resources with respect to the direct process fidelity estimation. Note however that the lower bound is expected to be an efficient estimator of the process fidelity for high-fidelity realizations in general. For the single-shot protocol we have a reward ${2}/(d\,\mathrm{Pr}(j_{n,m}))\mathrm{Tr}[\rho_n P_{j_{n,m}}]e^{j_{n,m}}_l$.

\begin{figure}[t]
\centering
\includegraphics[width=1.\linewidth]{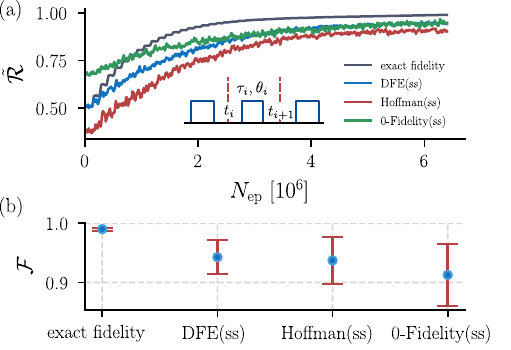}
\caption{
Learning and evaluation for indirect nuclear-spin control. (a) Smoothed episode reward $\tilde{\mathcal{R}}$ with a window of $10^3$ episodes. (b) Evaluation of three-pulse sequences implementing a Hadamard gate on the nuclear spin using control of the electronic spin only. The system parameters are $M_{zz}=-0.152~\mathrm{MHz}$, $M_{zx}=0.110~\mathrm{MHz}$, $\Omega=0.5~\mathrm{MHz}$, and $B_z=14.8~\mathrm{mT}$. The pulse parametrization is shown in the inset of (a); for each pulse, the agent selects two free-evolution times, the pulse duration, and phase within $[0,1]~\mu\mathrm{s}$, $[0,2]~\mu\mathrm{s}$, and $[-\pi,\pi]$, respectively. The learned sequences have total durations of about $5~\mu\mathrm{s}$, comparable to Refs.~\cite{PhysRevLett.124.220501,PhysRevApplied.12.064047,PhysRevLett.125.030501}.
}

\label{fig:fidelities_fig}
\end{figure}

\begin{table*}[!ht]
\centering
\small
\begin{tabularx}{0.8\textwidth}{lll}
\hline
Target & Description & Pulses \\
\hline
Hada4  & Single-qubit Hadamard gate on a $^{13}\mathrm{C}$ spin & $4$ cont. pulses \\
Hada10 & Single-qubit Hadamard gate on a $^{13}\mathrm{C}$ spin & $10$ cont. pulses \\
Bell16 & Bell-state preparation between two $^{13}\mathrm{C}$ spins & $16$ cont. pulses + DD \\
\hline
\end{tabularx}
\caption{Overview of target tasks. All tasks use finite-duration pulses, denoted by \textit{cont.} (for continuous). For Bell16, dynamical decoupling $($DD$)$ is implemented using instantaneous echo pulses interleaved with the finite-duration pulses.}
\label{tab:targets}
\end{table*}
In Fig.~\ref{fig:fidelities_fig} we show a first example that highlights the reward strategies at our disposal. We consider a simple example -- which however mimics closely~\cite{PhysRevLett.124.220501} --, where the full system consists simply of the electronic spin and a nuclear one. The physical parameters are reported in the figure. The RL agent is asked to learn a 3-pulse sequence to perform the Hadamard gate on the nuclear qubit having access to the single qubit measurements of the same qubit. Fig.~\ref{fig:fidelities_fig}(a) shows the learning curves under different reward schemes. Exact fidelity here stands for the case in which the reward is the process fidelity calculated exactly. The other curves are single-shot reward strategies employing direct fidelity estimation, Hofmann lower bound, and 0-fidelity to estimate the process fidelity. In Fig.~\ref{fig:fidelities_fig}(b), we see that the exact fidelity performs best, as expected, giving a trained agent that performs near optimally during evaluation. We see that the RL agents using single-shot strategies, while not performing as the exact fidelity one, still manage to learn high-fidelity solutions with the direct fidelity estimate strategy outperforming the other two during evaluation. From here on, when using single-shot reward we use the DFE estimator.

\subsection{Domain randomization for sim-to-real transfer}
Domain randomization (DR) is a widely used tool for sim-to-real transfer (see~\cite{zhao2020sim} and references therein), i.e., the deployment and adaptation of agents trained in a simulated environment into the real world. DR for sim-to-real transfer has been demonstrated to be effective in tasks such as robotics and autonomous driving. It consists of randomizing relevant physical parameters in the agent environment at each episode in such a way to produce robust policies that remain effective when transferred to the real world. The idea is that, given a simulator that reproduces the real dynamics as close as possible, the randomization of relevant parameters is enough to cover the real-world variability. DR can be used in zero-shot transfer, i.e., by deploying directly the trained agent in the real-world environment, or via fine-tuning by running further (limited) training in the real-world environment~\footnote{See also~\cite{reinschmidt2024reinforcement,peters2026deep} for sim-to-real transfer in the context of cold atoms experiments.}.

{In this work we use DR at two different levels. On the one hand, for emulating the completely mixed thermal state of the spin-bath within the gCCE, at each episode a different state-configuration of the spin-bath is uniformly sampled and the agent learns a strategy dealing with this uncertainty. This DR is always present in our simulations with a bath.
On the other hand,} we employ domain randomization to produce robust policies against changes in relevant parameters like the Rabi frequency of the mw pulses or the coupling constants of the register- and the bath nuclear spins.
Also in this second case, the randomization happens at each episode.
We then evaluate the robustness of our solutions via zero-shot deployment in (simulated) environments that are sampled from either the training distribution or are outliers.
Furthermore, we use the trained agents with DR as warm start for fine-tuning in sampled environments.

\section{Results}\label{sec:results}
We report now our main results showing the performance and adaptability of our RL framework. We do this by focusing on
three different examples whose targets are listed in Tab.~\ref{tab:targets}. These include both gate design on a single nuclear spin and state preparation of an entangled state between two nuclear spins via indirect control.

Generally, RL algorithms can be efficiently tailored to specific problems and circumstances by using information on the underlying physical system in which the agent operates.
{In the case of quantum control, apart from limiting the number of steps the agent is allowed to take, i.e. the length of the pulse sequences, also the action space can be chosen in a way that makes physical sense for the specific platform of interest. In our case, where the platform of interest is an NV-based nuclear spin register in diamond, the coupling between electron and nuclear spins that we consider is of the order of tens to hundreds of kilohertz. Hence, the pulse sequences should be built from time segments of the order of microseconds.}
For the specific tasks we consider here (cf. Tab.~\ref{tab:targets}), the action spaces comprise three actions corresponding to delay time, pulse time, and pulse phase. For the Hadamard tasks, they are restricted to $[0,1\rm{\mu s}]$ for the delay time,  $[0,1.6\rm{\mu s}]$ for the pulse time, and  $\phi\in [0,\pi]$ for the pulse phases. For the Bell state we use $[0,2\rm{\mu s}]$ (delay times), $[0,2\rm{\mu s}]$ (pulse times), and  $\phi\in [0,2\pi]$ (pulse phases). When the agent is required to learn the optimal amplitude of the magnetic field we set $B\in [0,50~\mathrm{mT}]$.
For the preparation of the Bell state between two $^{13}$C nuclear spins, a penalty on the actions is introduced in the form of $\propto a^4$ with action $a\in[-1,1]$ to boost exploration and avoid only extremal actions selection.

In the following, when showing the learning curves of our agents we always report the smoothed episode reward ($\tilde{\mathcal{R}}$) with a window size of $10^3$ episodes.

\subsection{Gate preparation: Hadamard gate }
Preparing a Hadamard gate on a nuclear spin via indirect control of the NV electronic spin is not an easy feat~\cite{PhysRevLett.124.220501,PhysRevApplied.12.064047,PhysRevLett.125.030501}. Here we tackle this problem with our RL agents.
{In order to consider the stability of the performance of RL control under different initial guesses for the number of pulses, we allow sequences with maximal length of either four (termed Hada4) or ten pulses (termed Hada10), see Tab.~\ref{tab:targets}.} Furthermore, we use either the exact process fidelity or the single-shot direct fidelity estimation as rewards. The former offers a high-performing strategy for training in simulations, while the latter can be employed on physical devices.
To understand the impact of the spin-bath on the training, the agents are trained either in an idealized system composed of the NV spin and the nuclear one or in the presence of a spin-bath via the first order gCCE (gCCE1 bath). We verify the truncation to gCCE1 by evaluating the trained policy in the exact spin-bath and find that gCCE1 captures the essential features of the open system dynamics on the timescale required for the Hadamard gate.

In Tab.~\ref{tab:agents} we summarize the agents trained for the gate design tasks.

\begin{table}[t]
\centering
\small
\begin{tabular}{lll}
\hline
Agent & spin-bath type & Reward type \\
\hline
A1-3 & no bath & exact fidelity \\
B1-3 & gCCE1 bath & exact fidelity \\
A1-3(ss) & no bath & single-shot fidelity \\
B1-3(ss) & gCCE1 bath & single-shot fidelity \\
\hline
\end{tabular}
\caption{Overview of agents used for the tasks Hada4 and Hada10. The three agents trained without the spin-bath are denoted by A, while the three agents trained with the spin-bath are denoted by B. }
\label{tab:agents}
\end{table}

\begin{figure*}[t]
\centering
\includegraphics[width=1.0\linewidth]{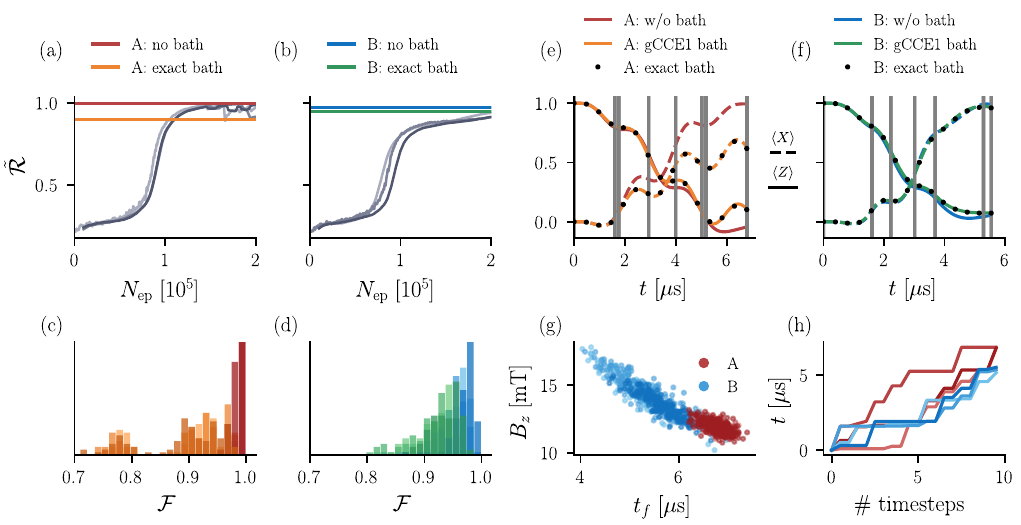}
\caption{\label{fig:Hada10}
RL agents learn pulse sequences robust to spin-bath effects for the Hada10 target. (a,b) Training curves for agents A1–3, trained without a bath (a), and B1–3, trained with a gCCE1 bath (b), using exact process fidelity as reward. Gray curves show training for independent runs; horizontal lines indicate mean fidelities evaluated without a bath and with the exact bath. (c,d) Fidelity distributions from stochastic policy evaluations. Bath-trained agents achieve higher exact-bath fidelities. (e,f) Representative dynamics from agents A and B, showing $\langle X\rangle$ (dashed) and $\langle Z\rangle$ (solid) without a bath, with gCCE1, and with the exact bath (dotted). Agent B remains robust to bath effects, with good agreement between gCCE1 and exact simulations. Final exact-bath fidelities are $F_A=78.6\%$ and $F_B=96.1\%$. (g) Final sequence duration $t_f$ and magnetic field $B_z$ from stochastic evaluations. Bath-trained agents learn shorter sequences and larger fields. (h) Cumulative sequence time over the 10 free-evolution intervals and 10 microwave pulses; plateaus indicate zero-duration steps. Agents B typically reach the target gate in less time.
}
\end{figure*}

\begin{table}[b]
\centering
\small
\begin{tabular}{llcc}
\hline
Target & Agent & No bath & Exact bath \\
\hline
Hada4 & A1-3 & 99.94\% & 88.27\% \\
 & B1-3 & 97.49\% & 96.34\% \\
 & A1-3(ss) & 96.95\% & 89.00\% \\
 & B1-3(ss) & 95.99\% & 93.44\% \\
Hada10 & A1-3 & 99.76\% & 89.59\% \\
 & B1-3 & 96.94\% & 94.47\% \\
\hline
\end{tabular}
\caption{Overview of mean fidelities. }
\label{tab:agents_fidelity}
\end{table}

\subsubsection{Learning in a structured spin-bath}\label{sec:structure-bath}

We start by discussing the learning process of three agents A (A1-3) and three agents B (B1-3) (cf. Table~\ref{tab:agents}) with the exact process fidelity as terminal reward. Figs.~\ref{fig:Hada10}(a) and (b) show the learning curves for both agents and the results of the evaluation of the agents on different environments for the Hada10 problem. First we see that all agents A and all agents B have similar learning curves, which proves reproducibility of the results and points toward low variance during learning. Second, we see how agents B perform better than agents A when evaluated in the environment accounting for the exact open system dynamics of the NV register. The same result can be seen in Figs.~\ref{fig:Hada10}(c) and (d) where the distribution of the process fidelity over several evaluations of the agents' stochastic policies is shown.
It is important to stress here that, while other black-box optimization methods furnish a single pulse sequence, RL with PPO produces a stochastic policy distribution. In evaluating the performance of the model we always consider the mean result and not the best case fidelity that can be seen in Figs.~\ref{fig:Hada10}(c) and (d) (cf. Tab.~\ref{tab:agents_fidelity}).
This is consistent with the need to deploy the agent on real devices. However, as far as a simulated optimization goes, the strategy with the best fidelity can be readily extracted.

In order to better understand how the solutions found by agents B outperform agents A in the presence of a spin-bath, we evaluate both policies 100 times and plot the learned magnetic field and the total gate times in the scatter plot of Fig.~\ref{fig:Hada10}(g). We find that agents B, facing the spin-bath modeled via gCCE1, learn to use shorter times and larger magnetic fields. While shorter times naturally mean less decoherence for the electron spin, the larger magnetic field increases the Zeeman splitting of the nuclear spins in the spin-bath effectively stiffening their quantization axis and suppressing their transverse spin dynamics.
Interestingly, Fig.~\ref{fig:Hada10}(h) shows that both agent groups learn that not all allowed pulses are needed and set thus the time duration of unneeded pulses to zero, resulting in sequences shorter than the available length.
Finally, Figs.~\ref{fig:Hada10}(e) and (f) show the dynamics of the target nuclear spin during the driving protocols created by agents A and agents B, respectively, with and without the spin-bath. While we see that the trajectories induced by agents A change substantially when a spin-bath is present, the trajectories induced by the sequence from agents B remain stable in the presence of the bath. Furthermore, Figs.~\ref{fig:Hada10}(e) and (f) justify a posteriori the use of the gCCE1 in the learning procedure showing the strong agreement with the dynamics of the exact bath.

Results for when the agent is restricted to use at most 4 pulses (Hada4 task) are shown in Appendix~\ref{app:Hada4} since they are qualitatively similar to the Hada10 task.

\begin{figure}[t]
\centering
\includegraphics[width=1\linewidth]{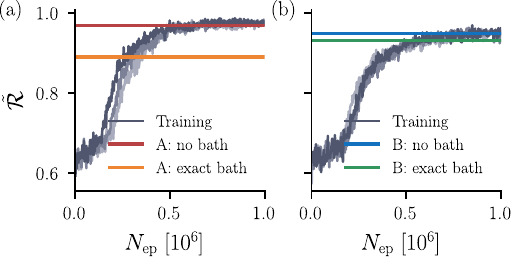}
\caption{\label{fig:learning_Hada4_ss}
Training with single-shot fidelity rewards for the Hada4 target. (a) Agents A1–3(ss), trained without a spin-bath. (b) Agents B1–3(ss), trained with the gCCE1 bath. Gray curves show independent training runs; horizontal lines indicate mean fidelities $\mathcal{F}$ evaluated without a bath and with the exact bath. Despite the noisy single-shot reward, bath-trained agents learn pulse sequences that are more robust to environmental effects.
}
\end{figure}
\subsubsection{Training with single-shot fidelity}
We now turn our attention to the performance of agents in group A1-3(ss) and B1-3(ss), i.e., to agents that employ a single-shot direct fidelity estimation reward strategy. While here we perform simulations akin to the ones shown in the previous section, we stress that a strategy based on single-shot (or even few shots) fidelity reward is amenable to be implemented directly on an experimental setup. Here we consider the case in which the agents are allotted 4 pulses to perform (Hada4 target). Fig.~\ref{fig:learning_Hada4_ss} shows once more that agents in group B learn to account for the effect of the spin-bath (cf.~Tab.~\ref{tab:agents_fidelity}). For the single-shot results for the Hada10 task we refer to Appendix~\ref{app:Hada4}.

It should be noted that, in the single-shot strategy case the agents require an order of magnitude more episodes to learn with respect to the case in which the exact process fidelity is used as reward. Furthermore, the variance of the policy also increases substantially as one can deduce from comparing the training curves in Figs.~\ref{fig:learning_Hada4_ss}(a) and (b) with the ones in Figs.~\ref{fig:Hada10}(a) and (b). Nevertheless, the comparison between different training curves in Fig.~\ref{fig:learning_Hada4_ss} again shows that the results are reproducible.

\begin{figure*}[t]
\centering
\includegraphics[width=1.0\linewidth]{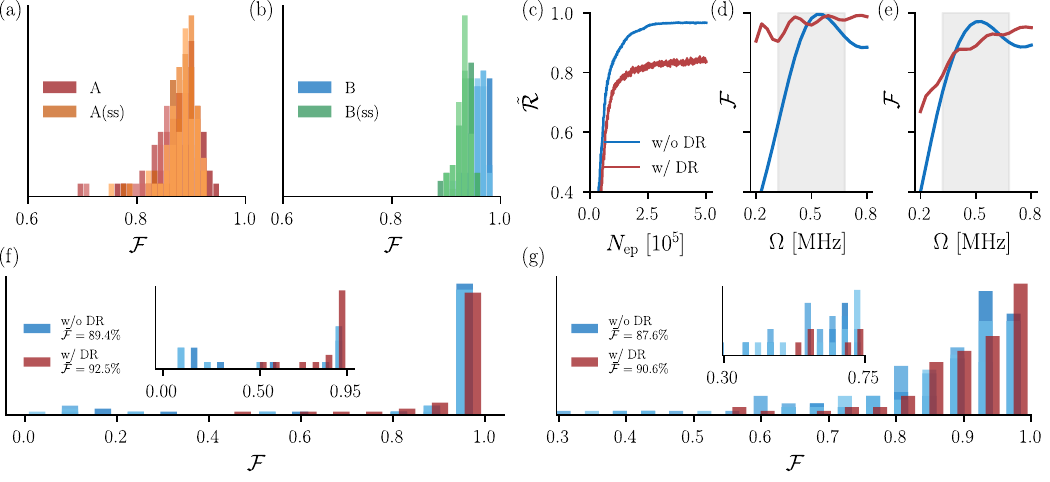}
\caption{\label{fig:DR}
Robustness and domain randomization for the Hada4 target. (a,b) Fidelity distributions under bath-spin position disorder, $\Delta x_j^{(k)}\sim\mathcal{N}(0,\sigma)$ with $\sigma=0.1~\mathrm{nm}$, and Rabi-frequency noise, $\Omega\sim\mathcal{N}(\mu_\Omega,\sigma_\Omega)$ with $\mu_\Omega=0.5~\mathrm{MHz}$ and $\sigma_\Omega=0.02~\mathrm{MHz}$. Agents trained with the gCCE1 bath, B1–3 and B1–3(ss), are more robust than agents trained without a bath, A1–3 and A1–3(ss). (c–e) Domain randomization of the Rabi frequency, using $\sigma_\Omega=0.2~\mathrm{MHz}$ during training. (c) Learning curves with and without domain randomization. (d,e) Evaluation over fixed $\Omega\in[0.2,0.8]~\mathrm{MHz}$ without a bath (d) and with the exact bath (e); the shaded region indicates the training range. (f,g) Position domain randomization under stronger evaluation disorder. (f) Bath-spin positions for agents B1–3 in the exact bath, with $\sigma_{\rm train}=0.4~\mathrm{nm}$ and $\sigma_{\rm eval}=1.0~\mathrm{nm}$, improving $\bar{\mathcal{F}}$ from $89.4\%$ to $92.5\%$. (g) Register $^{13}\mathrm{C}$ position disorder with $\sigma=0.05~\mathrm{nm}$ for agents A1–3 without a bath, improving $\bar{\mathcal{F}}$ from $87.6\%$ to $90.6\%$. Insets highlight the reduced occurrence of low-fidelity configurations.
}
\end{figure*}

\subsection{Domain randomization}
As we have already discussed, randomizing the physical parameters of an RL agent environment can result in more robust policies and aid in sim-to-real transfer. When using DR the agent is exposed to different environments, characterized by the occurrence of the sampled physical parameters that we randomized, at each episode.
We apply domain randomization to several relevant physical parameters. The first one is the Rabi frequency of the mw pulses. The Rabi frequency is randomly sampled from a normal distribution $\Omega\sim\mathcal{N}(\mu_\Omega, \sigma_\Omega)$ centered at $\mu_\Omega=0.5~\mathrm{MHz}$. We also consider randomizing the positions of the spins in the spin-bath of the NV register which is equivalent to randomizing the bath spin couplings to the register spins and the other spins in the bath. The position variations are sampled from $\Delta x_j^{(k)}\sim\mathcal{N}(0,\sigma)$ for $j=1,2,3$ and $k=1,\dots,N_{\rm bath}$. Finally, we also randomized the couplings of the register nuclear spin with the NV and with the bath spins. Here, the position variations are {sampled again from $\Delta x_j\sim\mathcal{N}(0,\sigma)$ for $j=1,2,3$}.

Firstly, we evaluate the robustness of the policies of our trained agents from the previous section with respect to Rabi frequency and spin-bath position fluctuations. Figs.~\ref{fig:DR}(a) and (b) show the stability of these policies for the Hada4 task. Notice that, agents in group B, that is the ones trained in the gCCE1 spin-bath, are more robust than agents in group A trained in the absence of the spin-bath.

We then apply DR with respect to the Rabi frequency, bath-spin positions, and register-spin position separately and show the results in Figs.~\ref{fig:DR}(c)-(g). Fig.~\ref{fig:DR}(c) shows the different training curves with and without DR. While the agent without DR learns faster, the one with DR yields higher fidelities when tested on different Rabi frequencies (see Figs.~\ref{fig:DR}(d) and (e)). In Fig.~\ref{fig:DR}(f), we show the fidelity histogram produced by evaluating both policies varying the Rabi frequencies and the spin-bath positions, while in Fig.~\ref{fig:DR}(g) we vary the Rabi frequency and the position of the register $^{13}$C spin. Overall, these figures show a clear increase in the performance of the domain-randomized agents once evaluated in environments that are sampled either from the same distribution used for training or even outside of it, cf. Figs.~\ref{fig:DR}(f) and (g), and also an increase in the robustness of the solutions, cf. Figs.~\ref{fig:DR}(c)-(e) and Figs.~\ref{fig:DR}(f) and (g)  insets.

\begin{figure*}[t]
\centering
\includegraphics[width=1.\linewidth]{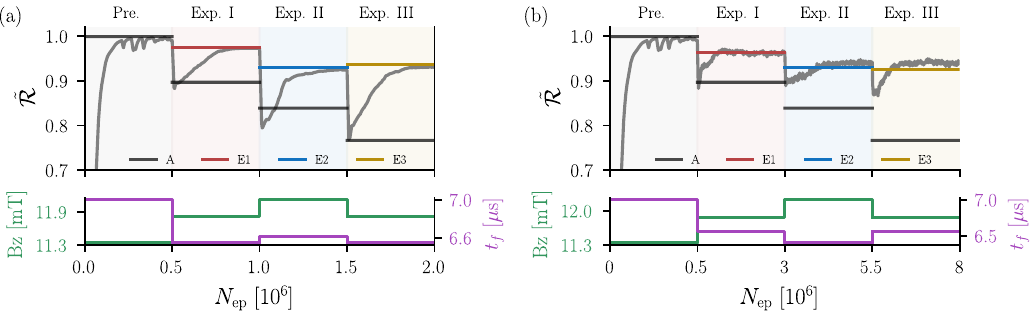}
\caption{\label{fig:pre-training_Hada4_combined}
Continued training after pre-training for the Hada4 target. A pre-trained agent is evaluated first and then used to initialize three subsequent training stages, E1–3, each corresponding to a new realization of bath-spin position disorder and Rabi-frequency noise. We stress that, the same pre-trained agent is used in E1-3 as a starting point.
Position shifts are sampled as $\Delta x_j^{(k)} \sim \mathcal{N}(0,\sigma)$ with $\sigma = 0.1~\mathrm{nm}$, and Rabi frequencies as $\Omega \sim \mathcal{N}(\mu_\Omega,\sigma_\Omega)$ with $\mu_\Omega = 0.5~\mathrm{MHz}$ and $\sigma_\Omega = 0.02~\mathrm{MHz}$.
(a) Continued training using the exact fidelity.
(b) Continued training using the single-shot fidelity estimator.
In both cases, continued training restores high fidelities and adapts the control parameters, leading to larger $B_z$ and shorter final sequence times $t_f$.
}
\end{figure*}

\subsection{Pre-training}

While domain randomization serves the purpose of enhancing the robustness of policies obtained in a simulated environment, zero-shot transfer to the real-world -- mimicked by the direct evaluation of the policies in a different environment as done in the previous section -- can still result in a decay in the performance. It is thus reasonable to fine-tune the agent's policy via extra limited training on the real-world environment when possible\footnote{That a pre-training strategy giving rise to a warm-start could offer speed-ups in the training of the agent was pointed out also in~\cite{PhysRevX.12.011059}.}.
Here we show what happens when using a trained agent as a warm start for extra training on an environment characterized by a different Rabi frequency or spin-bath positions with respect to the ones used during training.

First we consider an agent trained without any bath, i.e., agent A. After a pre-training phase (Pre.), we use the pre-trained agent and simulate three different experimental situations (Exp. agents E1-3). Each E1-3 corresponds to a new realization of bath-spin
positions and Rabi-frequency. Note that, for each E1-3 the initial agent corresponds to the pre-trained agent A.
In Fig.~\ref{fig:pre-training_Hada4_combined}, we show the results of this analysis. We see that, after a pre-training phase, the agent is able to adapt to each new environment and once again learns higher magnetic fields and shorter duration pulse sequences than in the pre-training phase.
In particular, Fig.~\ref{fig:pre-training_Hada4_combined}(b) shows how this is effective when transitioning from a pre-training phase to a fine-tuning using single-shot fidelity reward. In Appendix~\ref{app:Hada4} the same analysis is carried out for the Hada10 task showing that, while without pre-training agents B(ss) are not able to learn shorter pulse sequences, once pre-training is introduced agents B(ss) go on learning higher magnetic fields and shorter duration pulse sequences.

\begin{figure}[!b]
\centering
\includegraphics[width=1\linewidth]{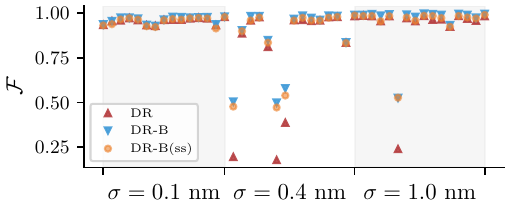}
\caption{\label{fig:pretraining_DR_manyagents_A}
Pre-training with DR and continued training with DR of bath-spin positions for the Hada4 target. For each variance $\sigma$, $15$ randomized bath realizations are evaluated. The pre-trained DR agent is compared with continued training using either the exact fidelity, DR-B, or the single-shot estimator, DR-B(ss). Continued training recovers high fidelities for most realizations, after $N=5\times10^5$ shots for DR-B and $N=2.5\times10^6$ shots for DR-B(ss).
}
\end{figure}

Finally, Fig.~\ref{fig:pretraining_DR_manyagents_A} shows an extensive evaluation analysis for a pre-trained agent with bath-spin position DR, $\Delta x_j^{(k)}\sim\mathcal{N}(0,\sigma)$ with $\sigma=0.4~\mathrm{nm}$, and Rabi-frequency DR, $\Omega\sim\mathcal{N}(\mu_\Omega,\sigma_\Omega)$ with $\mu_\Omega=0.5~\mathrm{MHz}$ and $\sigma_\Omega=0.02~\mathrm{MHz}$. The pre-trained agent is first evaluated, then fine-tuned by continuing learning with either exact fidelity reward or single-shot DFE reward, and finally evaluated again on 45 different environment realizations. The latter are characterized by different variances in the position distribution of the bath-spins.
We see that the fine-tuning restores high fidelities for most bath realizations, with improvements observed with both reward types.

{Both cases shown in this section can be seen as blueprints of what could be done on a real device. The first one highlights that fine-tuning following a generic pre-training phase in which only the information about the register system is available still produces good results although neglecting altogether the spin-bath dynamics. The second case is the one closer to the sim-to-real transfer practice, where we put together all the tools considered thus far and pre-trained an agent on a `realistic' model of a nuclear register, accounting for variability in the physical parameters via DR, and then fine-tuned the agent on a different environment. Importantly, our results also point toward the fact that the fine-tuning can be performed with a limited number of episodes after the pre-training on a simulated environment.}

\subsection{State preparation: Bell state }
We now turn our attention to a state preparation task via indirect control. In particular, we use our RL agents to find a pulse sequence to prepare a Bell state between two $^{13}$C nuclear spins. Here, we use the NV-center as a bus for generating the entanglement, as the direct dipolar coupling between the nuclear spins is very small. Furthermore, because the second register $^{13}$C spin is located further away the pulse sequence becomes longer and the control more difficult. To deal with this problem, we use the exact state fidelity as a reward and we do not task the agent with also learning the optimal external magnetic field. Additionally, we also interleave finite duration pulses with instantaneous echo pulses that are placed in the middle of the free evolution intervals as done in~\cite{PhysRevB.110.035407}.
As shown in Tab.~\ref{tab:benchmarks}, while RL is able to find high fidelity pulse sequences for this task the same is not true for other black-box optimization methods even in the absence of a spin-bath.

\begin{figure*}[t]
\centering
\includegraphics[width=0.8\linewidth]{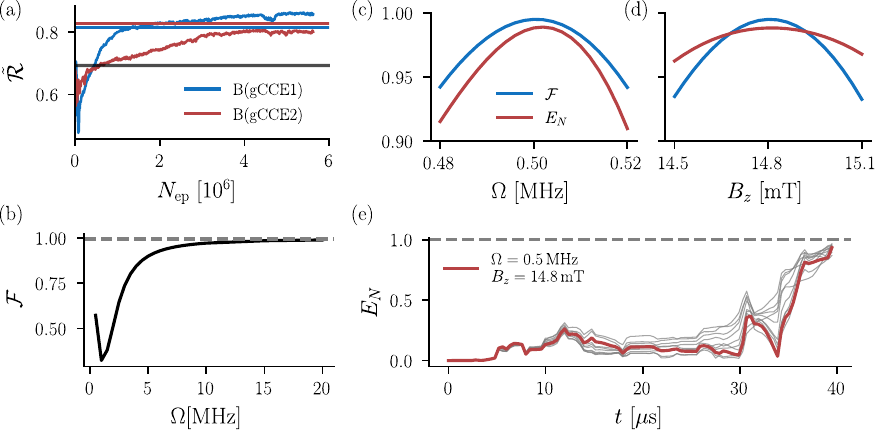}
\caption{\label{fig:Bell16}
Training and robustness for the Bell16 target. (a) Continued training after pre-training without a spin-bath (agent A, gray), using either gCCE1 or gCCE2 bath models. The reduced exact-bath fidelity for gCCE1 shows that higher-order bath correlations become important for the longer Bell-state preparation time, making gCCE2 more appropriate. (b) Effect of finite DD-pulse durations. Fidelities above $90\%$ are obtained for $\Omega\gtrsim5~\mathrm{MHz}$; the horizontal line marks the instantaneous-pulse limit, $F_{\mathrm{inst}}=99.49\%$ (Tab.~\ref{tab:agents_Bell_fidelity}). (c,d) Robustness of the pulse sequence learned by agent A to variations in the Rabi frequency $\Omega$ (c) and magnetic field $B_z$ (d). (e) Logarithmic negativity $E_N$ during the Bell16 sequence for $\Omega=0.5~\mathrm{MHz}$ and $B_z=14.8~\mathrm{mT}$; gray curves show $10$ samples with $\sigma_\Omega=0.02~\mathrm{MHz}$ and $\sigma_{B_z}=0.3~\mathrm{mT}$.
}
\end{figure*}

\begin{table}[!b]
\centering
\small
\begin{tabular}{llccc}
\hline
Target & Agent & RL & CMA-ES & BO \\
\hline
Hada4 & A & 99.98\% & 99.98\% & 98.73\% \\
 & B & 97.00\% & 96.60\% & 93.21\% \\
 & A(ss) & 97.70\% & 99.08\% & 99.30\% \\
 & B(ss) & 93.66\% & 93.44\% & 88.31\% \\
Hada10 & A & 99.79\% & 99.98\% & 94.75\% \\
 & B & 95.60\% & 94.50\% & 80.71\% \\
Bell16 & A & 99.49\% & 70.96\% & 70.78\% \\
\hline
\end{tabular}
\caption{Comparison of the best performances of reinforcement learning agents (RL) with the best results from closed-loop black-box optimization techniques: covariance matrix adaptation evolution strategy (CMA-ES), and Bayesian optimization (BO). The same number of episodes was allowed for the benchmarks (even though often not all episodes were used because the algorithms got stuck in local minima). For the benchmarks the objective has been chosen as the average of $N=20$ episodes with different initial states of the spin-bath for B and $N=500$ episodes for A(ss) and B(ss) corresponding to 500 shots (see main text). For more information about the convergence see Figs.~\ref{fig:optimization}(a) and (b). }
\label{tab:benchmarks}
\end{table}

We proceed as in the previous section and consider a pre-trained agent that is further fine-tuned. In particular, agent A is pre-trained with a Rabi frequency $\Omega=0.5~\mathrm{MHz}$ and magnetic field $B_z=14.8~\mathrm{mT}$ in the absence of the spin-bath. Upon evaluation it reaches a state fidelity of $99.5\%$ and a logarithmic negativity $E_N=0.988$ while its performance decreases substantially when evaluated in the presence of the spin-bath ($\mathcal{F}\approx 70\% $).
After the pre-training phase, we continue the agent training by including either the gCCE1 approximation to the spin-bath (agent B(gCCE1)) or the gCCE2 approximation (agent B(gCCE2)).
Fig.~\ref{fig:Bell16}(a) shows the learning curves and the results of the evaluation while Table~\ref{tab:agents_Bell_fidelity} reports the state fidelity upon evaluation. We see that continuing the training by including the spin-bath via gCCE allows us to reach state fidelities greater than 80$\%$ when evaluating the policies in the exact spin-bath. It is interesting to note that the exact bath state fidelity obtained from the agent trained within the gCCE2 approximation of the spin-bath is higher than the one obtained from the agent trained with the gCCE1 spin-bath. This points to the fact that, for this task that entails long duration pulse sequences, higher order contributions to the spin-bath correlations have to be taken into account.

Figs.~\ref{fig:Bell16}(c)-(e) highlight the robustness of the solution of agent A to variations in the Rabi frequency and in the external magnetic field magnitude as well as the generation of entanglement during the pulse sequences dictated by the agent policy.
{We see that also the entanglement generation dynamics, and the overall negativity, are quite robust against variations in the Rabi frequency and external magnetic field.}

Finally, it should be noted that here we have assumed the dynamical decoupling (DD) echo $\pi$-pulses to be instantaneous. This is a common approximation in the DD literature and it is also customary to use the higher allowed Rabi frequency for DD pulses while maintaining lower Rabi frequencies for coherent control in actual experiments~\cite{liu2013noise}.
To obtain a better understanding of the necessary Rabi frequencies needed for an effective instantaneous pulse, we compare the Rabi frequency and the coupling to the respective $^{13}$C spin. The dipolar coupling to the closest spin is of the order of $0.1$ MHz. Hence rotating the electron spin with tens of MHz, the nuclear spins cannot follow the trajectory of the electron spin fast enough.  We
show the fidelity of the gate by changing the Rabi frequency of the DD pulses in the found solution in  Fig.~\ref{fig:Bell16}(b). We see that Rabi frequencies $\Omega\geq 10$~MHz are already enough to recover the high fidelity of the instantaneous pulse limit.

\begin{table}[b]
\centering
\small
\begin{tabular}{llcc}
\hline
Target & Agent & No bath & Exact bath \\
\hline
Bell16 & A & 99.49\% & 69.30\% \\
 & B(gCCE1) & 80.83\% & 81.50\% \\
 & B(gCCE2) & 92.51\% & 82.77\% \\

\hline
\end{tabular}
\caption{Overview of fidelities for the Bell16 target. Agent B(gCCE1) trained with gCCE1 bath, agent B(gCCE2) trained with gCCE2 bath. Testing using the exact thermal bath. }
\label{tab:agents_Bell_fidelity}
\end{table}

\section{Benchmarking: black-box control techniques}\label{sec:bench}
Our model-free RL agents act as a black-box optimization method suitable, when considering single-shot rewards, for the direct training on a real device.
As we have discussed, this allows one to circumvent model bias.
However, model-free RL is not the only black-box optimization tool. Indeed, different approaches have been proposed for exactly this kind of scenario. We thus benchmark our agents against black-box optimization methods suitable for noisy objective functions. In particular, we consider Bayesian optimization (BO) and evolution strategies\footnote{Also simultaneous perturbation stochastic approximation (SPSA) was used but did not yield comparable results.}.
While we give more details on these methods in Appendix~\ref{app:blackbox}, it is important to notice that these black-box methods are equipped with the same physics-informed constraints that we used for shaping the action space of our agents.

Table~\ref{tab:benchmarks} shows the results obtained in comparison with our RL agents. When comparing with the agents in group B we use the black-box optimization methods in the same gCCE1 spin-bath environment with objective function given by the average of the process fidelity over $N=20$ different initial states of the spin-bath. This serves to emulate the completely mixed state of the spin-bath at evaluation. Notice that the RL agents do not explicitly benefit from such an average but learn with a changing initial state at each episode. This creates a bias in favor of the black-box benchmarks which nonetheless underperform in comparison.
Furthermore, we should stress that the black-box benchmarks used here do not work when using single-shot direct fidelity estimation as the objective function\footnote{A single-shot adaptation of BO has been proposed in~\cite{PRXQuantum.1.020322}. While we do not explore this method, we notice that for the sake of gate design the method is bound to require a large number of surrogates.}. Thus, for groups A(ss) and B(ss) the objective function of the benchmarks is given by averaging single-shot results over 500 episodes. This renders the comparison with the agents using the single-shot reward strategy all the more significant.

\begin{figure}[t]
\centering
\includegraphics[width=1.0\linewidth]{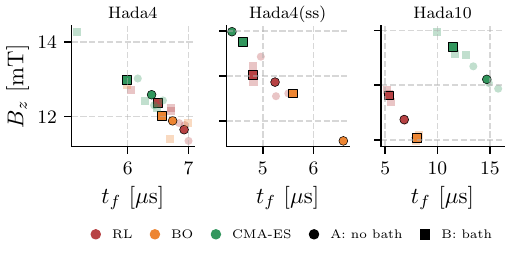}
\caption{\label{fig:bbcomparison}
Comparison of RL with Bayesian optimization (BO) and CMA-ES for the Hada4, Hada4(ss), and Hada10 targets. Final sequence duration $t_f$ and magnetic field $B_z$ are shown for optimization without (A) and with (B) a spin-bath. Light markers denote individual runs and opaque markers the corresponding means. The methods converge to distinct parameter regimes while achieving high-fidelity pulse sequences. See also Tab.~\ref{tab:benchmarks}.
}
\end{figure}

In Fig.~\ref{fig:bbcomparison} we show how the different methods compare not only in the final fidelity but also in terms of pulse sequence duration and learned magnetic field for the Hada4 and Hada10 targets. We see that the different methods seem to explore distinct parameter regimes, reflecting different optimization strategies for reaching high-fidelity pulse sequences. This is supported by the analysis of the time spent in different parts of the Bloch sphere during the pulse sequences that can be found in Appendix~\ref{app:BlochSphere}.
That the RL agents can discover qualitatively different high-fidelity protocols and reach parameter space regions not explored by other quantum optimization strategies has been highlighted in the literature~\cite{preti2024hybrid,zhang2019does,bukov2018reinforcement,khalid2021reinforcement} and represents in itself an intrinsic value potentially offsetting the long computational times required during simulations.

\section{Conclusions}\label{sec:conclusions}
{In this work we have investigated reinforcement learning for quantum optimal control in a challenging setup: learning in the presence of a structured spin-bath and by indirectly controlling a nuclear spin register via mw pulses on a central electronic spin. While our study is theoretical and employed detailed simulations, we have shown the learning abilities of the agents when subject to uncertainty in the physical parameters and when offered realistic single-shot Pauli measurement outcomes as reward.

Our results show that, also in this challenging situation, RL agents can be engineered that find high-fidelity solutions for both gate design and state preparation. Moreover, these solutions are comparable to, and often better than, the solutions found by other black-box optimization algorithms.

Crucially, using domain randomization during training, combined with fine-tuning at deployment and the use of single-shot fidelity rewards, we are able to lay down a blueprint for the experimental use of our agents. First, domain randomization accounts for the variability of physical parameters of a real-world scenario in simulation. This allows us to train agents in a simulated system+spin-bath that resemble as much as possible the experiment of interest -- if only information about the register is available one can also just train an agent in the absence of a spin-bath.
The agents trained this way use exact fidelity as rewards and learn policies robust against parameter fluctuations.
Such agents can then be fine-tuned in the real-world system with single-shot fidelity rewards. These rewards require simple Pauli measurements and separable state preparation and each measurement result can be used for the training. In our work we mimicked this real-world fine-tuning by sampling different environments, also outside the parameter range the agents see during pre-training, and continuing the training for a limited number of episodes using single-shot fidelity reward. In this way, a clear pipeline for the ready use of our agents in real-world experimental situations is laid out.

To conclude, we have shown that RL agents can learn to cope with the presence of a non-Markovian spin-bath and, via the use of sim-to-real transfer techniques developed in the RL literature, find solutions robust to uncertainty in the theoretical modeling of the environment. This shows the potential for applications in real solid-state qubit devices -- even beyond NV centers -- for quantum optimal control tasks.
}

\section*{Acknowledgments}
This project was made possible by the DLR Quantum Computing Initiative and the Federal Ministry for Research, Technology and Space;
qci.dlr.de/projects/qlearning and the Science Sphere concept of Ulm University. D.H., M.R., J.A. and D.M.  acknowledge funding through
the IQST and BMFTR (QSENS-QMAT).

\bibliographystyle{unsrt}
\bibliography{refs.bib}
\onecolumn
\newpage
\appendix

\section{Reinforcement learning in partially observable environments}\label{app:RL}
We briefly discuss here the concept of a partially observable environment in reinforcement learning. We refer the reader to the existing literature on reinforcement learning for more details and theoretical treatment.

As already discussed in the main text, in RL algorithms, an agent selects actions which modify the state of its environment and acquires rewards depending on the effectiveness of the action taken. The objective of the agent is to maximize the expected return, i.e., the statistical expectation of the cumulative sum of rewards received by the agent. The learning proceeds in episodes divided into time steps: At each time step the agent selects an action and (can) receive a
reward. But, while in many incarnations of RL the agent is given the full information about the state of the environment -- and this drives the learning -- in applications of RL to physical systems, from robotics to quantum physics, it is often the case that full observability of the environment is not an option. In these cases the agent has access to observations containing partial information on the state of the environment. This is the case of interest for us where the agent's environment becomes minimally observable with the agent having only the information about the learning time elapsed during the episode.

Formally, we consider a partially observable Markov decision process (POMDP). POMDPs are described by a 6-tuple $(\mathcal{S}, \mathcal{A},\mathcal{P}, \mathcal{R},\Omega,\mathcal{O})$. At each time step in an episode the agent partially observes a state $s_t \in \mathcal{S}$ via an observation $o_t\in \Omega$ -- assigned according to $o_t\sim \mathcal{O}(o_t|s_t)$ -- and chooses an action $a_t \in \mathcal{A}$. The agent's action affects the environment causing a transition to a new state determined by the (unknown to the agent) Markov transition function $s_{t+1}\sim \mathcal{P}(s_{t+1}|s_t,a_t)$ and, at each step, the agent receives a reward $r_t \sim \mathcal{R}(o_t,a_t,o_{t+1})$. At the end of an episode, that for our problem has a finite horizon $T$, we are left with a trajectory $\tau = \{o_0,a_0,r_1; o_1,a_1,r_2;\dots o_{T-1},a_{T-1},r_{T},s_{T}\}$ for which a return $G_{t}(\tau) = \sum_{k=t+1}^{T}\gamma^{k-t-1} r_k$ can be defined~\footnote{Here $\gamma\in [0,1)$ is a discount factor.}. The agent's objective is then the maximization of the total expected return
\[
J = \mathbb{E}_{\tau\sim P_{\pi}}\left[G_0(\tau)|o = o_0 \right].
\]
Here the average is taken over the probability of the trajectory to happen according to the dynamics of the system when actions are selected according to the agent policy $\pi(a_t|o_t)$.

In modern deep RL the policy is a parametrized policy $\pi_{\theta}$ represented by a deep neural network. Policy gradient RL algorithms try to directly learn an optimal policy maximizing $J(\theta)$ via gradient ascent (and generalizations thereof). Another family of RL algorithms, called value-function based methods, try to learn the parametrized state-action value function $Q^\pi_\phi(s,a)$
that is the expected return if starting in a certain state-action pair. This function is also represented via deep neural networks in modern deep RL.
Finally, actor-critic methods, like the PPO algorithm used in this work, combine policy gradient with value functions: an actor network learns the policy while a critic network learns a value function.

\section{Single-shot rewards: detailed derivation}\label{app:rewards}
Here we detail the derivation of the single-shot fidelities used throughout the main text.

\subsection{State fidelity}\label{app:statefid}
We start by considering state engineering of a target pure state with density matrix $\rho$. The state fidelity between this target and a (possibly mixed) state $\tilde{\rho}$ is given by
\begin{equation}
    F=\mathrm{Tr}(\rho\tilde{\rho})= \sum_{i=1}^{d^2} \mathrm{Tr}[\rho P_i] \mathrm{Tr}[\tilde{\rho} P_i]/d
\end{equation}
where $d=2^n$, $n$ is the number of qubits, and $P_i$ are the Pauli operators with $\mathrm{Tr}[P_i P_j]= d \delta_{ij}$.

Following~\cite{PhysRevX.12.011059}, the state fidelity can be written as
\begin{align}
    F \sim \frac{1}{N}\sum_{n}R^{(n)},
\end{align}
where $R^{(n)}=\pm 1$ are the binary outcomes that we obtain if we assume to be able to measure the dichotomic POVMs $\{ \ket{\psi}\bra{\psi}, 1-\ket{\psi}\bra{\psi}\}$, where $\ket{\psi}$ is the target state. In this case, the single-shot rewards for the RL agent are just the binary outcomes from the dichotomic measurements.
However, realistically the target state will not be a separable state and we will not have access to the dichotomic POVM projecting on it. Thus we need to follow a different route as exemplified in Ref.~\cite{PhysRevX.12.011059} appendix D (b).
We rewrite the fidelity as
\begin{align}
    F &= \sum_{i=1}^{d^2} \mathrm{Pr}(i) \mathrm{Tr}[\tilde{\rho} P_i]/\mathrm{Tr}[\rho P_i]\\ \nonumber
    & =\mathbb{E}_{i\sim Pr}\left[\mathrm{Tr}[\tilde{\rho} P_i]/\mathrm{Tr}[\rho P_i]\right]\\ \nonumber
    & =\frac{1}{d}\mathbb{E}_{i\sim Pr}\left[\mathrm{Tr}[\tilde{\rho} P_i]\mathrm{Tr}[\rho P_i]/\mathrm{Pr}(i)\right]
\end{align}
where $\mathrm{Pr}(i)=\mathrm{Tr}[\rho P_i]^2/d$ is the relevance distribution telling us how relevant a certain Pauli operator is in determining the state fidelity (in terms of their index $i$).

Note that, in the last expression we have still the quantity $\langle P_i\rangle_{\tilde{\rho}} = \mathrm{Tr}[\tilde{\rho} P_i]$
that needs to be estimated experimentally. This is an expectation value over the ``experimental'' state $\tilde{\rho}$, so we can rewrite the final expression formally as
\begin{equation}
    F= \frac{1}{d}\mathbb{E}_{\tilde{\rho}}\mathbb{E}_{i\sim Pr}\left[P_i \mathrm{Tr}[\rho P_i]/\mathrm{Pr}(i)\right].
\end{equation}
This comes from the eigenvalue decomposition of a Pauli operator $P_i= \sum_j e^{(i)}_j \ket{j}\bra{j}$.
Indeed, we have
\begin{equation}
    \mathrm{Tr}[\tilde{\rho} P_i]=\sum_j e^{(i)}_j \bra{j}\tilde{\rho} \ket{j}= \sum_j e^{(i)}_j p_{\tilde{\rho}}(j).
\end{equation}

Finally we can estimate the last expression via $\mathrm{Tr}[\tilde{\rho} P_i]\sim 1/M\,\, \sum_m e^{(i)}_m$ given the results of Pauli measurements on the state $\tilde{\rho}$ (which are the binary-valued eigenvalues). Thus all together

\begin{equation}\label{eq:appstfid}
    F\sim \frac{1}{d\, N\, M}\sum_n\sum_m \frac{1}{\mathrm{Pr}(i_n)}\mathrm{Tr}[\rho P_{i_n}]e^{(i_n)}_m.
\end{equation}

We see here that a viable single-shot reward for the RL agent is given by $1/d \frac{1}{\mathrm{Pr}(i_n)}\mathrm{Tr}[\rho P_{i_n}]e^{(i_n)}_m$ where the binary eigenvalues $e^{(i_n)}_m$ appearing result from the measurement of Pauli operators. Indeed, we have just seen that this expression, on average, returns a faithful estimator of the state fidelity. Note moreover that eq.~\eqref{eq:appstfid} allows us to also consider more than a single shot by setting $N$ and $M$ different from one. This allows us to use a better but still feasible state fidelity estimator as reward.

The complete protocol for the fidelity estimation is then:
\begin{itemize}
    \item select $N$ indices $i_n$ according to the relevance distribution, this sets the measurement Pauli (measurement settings)
    \item for each index $i_n$ measure the measurement Pauli $M$ times
\end{itemize}

\textbf{1. Note that} for few qubits it could be more efficient to consider all the Pauli operators that have a non-negligible weight according to the relevance distribution without sampling them, thus avoiding one of the arithmetic averages. However, this is then a $\text{number of Pauli operators}\times M$ shots protocol.

\textbf{2. Note that,} in~\cite{PhysRevLett.106.230501} a prescription for the number of shots that depends on the measurement is defined. Here we have not followed this scheme since it is not relevant for the single-shot reward. However, when considering more shots it could be convenient to look at a variable number of shots-per-Pauli.

\subsection{Gate engineering: Full direct fidelity estimation}\label{app:dfe}
We use the results in~\cite{PhysRevLett.106.230501}. The process fidelity between a unitary gate and a quantum channel is expressed as $F=\mathrm{Tr}[U^\dag \Lambda]/d^2$ where here $U$ and $\Lambda$ represent the superoperator matrices of the target gate and the experimental channel respectively.
The fidelity can be rewritten as
\begin{equation}\label{DFE1a}
    F=\frac{1}{d^2} \sum_{k,k'=1}^{d^2}\chi_U(k,k')\chi_\Lambda(k,k')
\end{equation}
where $\chi_U(k,k')= \mathrm{Tr}[P_k U(P_{k'})]/d$ and, analogously, for $\Lambda$. Here we consider $\{P_k\}$ to be the Pauli operators basis.
We can rewrite eq.~\eqref{DFE1a} as
\begin{align}\label{DFE2a}
    F &= \frac{1}{d^2} \sum_{k,k'=1}^{d^2}[\chi_U(k,k')]^2\chi_\Lambda(k,k')/\chi_U(k,k')\\ \nonumber
    & =\sum_{k,k'=1}^{d^2}\mathrm{Pr}(k,k')\chi_\Lambda(k,k')/\chi_U(k,k')\\ \nonumber
    & =\mathbb{E}_{i\sim Pr}[\chi_\Lambda(k_i,k_i')/\chi_U(k_i,k_i')],
\end{align}
i.e. as the expectation value over the probability $\mathrm{Pr}(k,k')$. This can be estimated via
\begin{equation}\label{DFE21a}
    F\sim \frac{1}{N}\sum_{i=1}^N[\chi_\Lambda(k_i,k_i')/\chi_U(k_i,k_i')]
\end{equation}

Note that the Pauli operators are not density matrices, i.e., valid states. Thus, in order to get close to an experimental protocol we need to expand the Pauli operators. We choose to use a separable basis of their eigenvectors since these are easy-to-prepare states. We thus have
\begin{align}\label{DFE3a}
    F &=\sum_{k,k'=1}^{d^2}\frac{\mathrm{Pr}(k,k')}{\chi_U(k,k')} \chi_\Lambda(k,k')\\ \nonumber
    & =\sum_{k,k'=1}^{d^2}\frac{\mathrm{Pr}(k,k')}{\chi_U(k,k')} \sum_{m=1}^{d}\frac{e^{(k')}_m}{d} \mathrm{Tr}[P_k\Lambda(\ket{v^{k'}_m}\bra{ v^{k'}_m})]\\ \nonumber
    & =\mathbb{E}_{i\sim Pr}\mathbb{E}_{m\sim \mathrm{uniform}}[e_m^{k'_i}\frac{\mathrm{Tr}[P_{k_i}\Lambda(\ket{v^{k_i'}_m}\bra{ v^{k_i'}_m})]}{\chi_U(k_i,k_i')}].
\end{align}
This can be estimated via
\begin{equation}\label{DFE31a}
    F\sim\frac{1}{N\,M}\sum_{i=1}^N\sum_{m=1}^M [e_m^{k'_i}\frac{\mathrm{Tr}[P_{k_i}\Lambda(\ket{v^{k_i'}_m}\bra{ v^{k_i'}_m})]}{\chi_U(k_i,k_i')}].
\end{equation}

Finally, in order to get a binary reward, and also for mimicking an experiment, we need to estimate the expectation values $\mathrm{Tr}[P_{k_i}\Lambda(\ket{v^{k_i'}_m}\bra{v^{k_i'}_m})]$ according to the statistics of the final experimental state $\Lambda(\ket{v^{k_i'}_m}\bra{v^{k_i'}_m})$.
We thus get
\begin{align}\label{DFE4a}
    F &=\sum_{k,k'=1}^{d^2}\frac{\mathrm{Pr}(k,k')}{\chi_U(k,k')} \sum_{m=1}^{d}\frac{e^{(k')}_m}{d} \sum_{m'=1}^{d}e^{k}_{m'}\mathrm{Tr}[\ket{v^{k}_{m'}}\bra{v^{k}_{m'}}\Lambda(\ket{v^{k'}_m}\bra{v^{k'}_{m}})]\\ \nonumber
    & =\mathbb{E}_{i\sim Pr}\mathbb{E}_{m\sim \mathrm{uniform}}\mathbb{E}_{m'\sim \mathrm{Tr}[\ket{v^{k_i}_{m'}}\bra{v^{k_i}_{m'}}\Lambda(\ket{v^{k_i'}_m}\bra{ v^{k_i'}_m})]}[\frac{e_{m}^{k'_i}e_{m'}^{k_i}}{\chi_U(k_i,k_i')}],
\end{align}
which can be estimated via
\begin{equation}\label{DFE41a}
    F\sim \frac{1}{N\,M\, L}\sum_{i=1}^N\sum_{m=1}^M\sum_{m'=1}^L [\frac{e_{m}^{k'_i}e_{m'}^{k_i}}{\chi_U(k_i,k_i')}].
\end{equation}

From this last expression we see that the reward $[\frac{e_{m}^{k'_i}e_{m'}^{k_i}}{\chi_U(k_i,k_i')}]$ can be used in the RL algorithm where $e_{m'}^{k_i}$ is the binary outcome of a local measurement, $e_{m}^{k'_i}$ is related to the initial separable state of the current experiment, and the rest are functions of the target gate and the choice of measurement and preparation.

The complete protocol for the DFE is then:
\begin{itemize}
    \item select $N$ pairs of indices $(k_i,k'_i)$ according to the relevance distribution, this sets the measurement Pauli (measurement settings) and the input Pauli
    \item for each index $i$ sample according to the uniform distribution $M$ separable eigenvectors of the input Pauli as input states
    \item finally measure the measurement Pauli for each of the input states. In the simulation this is accomplished via sampling from the state obtained by applying the experimental channel to the initial states and sampling according to this state $L$ eigenvalues of the measurement Pauli
    \item for single-shot reward in the RL algorithm we set $N=M=L=1$ at each iteration
\end{itemize}

\textbf{Note that,} in~\cite{PhysRevLett.106.230501} a prescription for the number of shots that depends on the measurement and input states is defined. Here we have not followed this scheme since it is not relevant for the single-shot reward. However, when considering more shots it could be ideal to look at a variable number of shots-per-Pauli-pairs.

\subsection{Gate engineering: Hofmann classical fidelity bounds}\label{app:hoffman}
A lower bound to the process fidelity is given in~\cite{PhysRevLett.94.160504} in terms of two classical fidelities. The bound reads
\begin{equation}
    F\geq F_1+F_2-1
\end{equation}
where $F_{i}=\frac{1}{d}\sum_{n=1}^{d} F_{i,n}$ with $i\in\{1,2\}$ and
\begin{equation}
    F_{i,n} = \mathrm{Tr}(U \ket{n}_i\bra{n}U^\dag\,\,V\ket{n}_i\bra{n}V^\dag).
\end{equation}
In this last expression the states $\ket{n}_i$ are chosen as the computational basis for $F_1$ and the $x$-basis for $F_2$.

Let us start by considering the set of $2^{n+1}$ states $\{\ket{z_n} ,\ket{x_n}\}$. We see immediately that the lower bound on the fidelity can be written as
\begin{align}
    F_1+F_2-1 &= \frac{2}{2d}\sum_{m=1}^{2d}F_m -1\\ \nonumber
    & = \mathbb{E}_{n\sim U(2d)}[2F_n-1]\sim \frac{1}{N\,M}\sum_{nm}R^{(n)}_m
\end{align}
where the first average is over the uniform distribution of indices of the states in  $\{\ket{z_n} ,\ket{x_n}\}$ and $R^{(n)}_m=\pm 1$ are the binary outcomes that we obtain if we assume to be able to measure the dichotomic POVMs $\{U \ket{n}\bra{n}U^\dag, 1-U \ket{n}\bra{n}U^\dag\}$ for all $n$. In this case, the single-shot rewards for the RL agent are just the binary outcomes from the dichotomic measurements.

In the more realistic case in which only local measurements of Pauli operators are available (i.e., we cannot measure in the basis containing the final states), we can proceed as before and expand the lower bound expression further. In particular we have:
\begin{align}
    F_1+F_2 &= \mathbb{E}_{n\sim U(2d)}[2F_n-1]\\ \nonumber
    &= \frac{1}{d}\mathbb{E}_{n\sim U(2d)}\mathbb{E}_{j\sim Pr^{(n)}}\mathbb{E}_{\psi_n}[\frac{2}{Pr^{(n)}(j)}\mathrm{Tr}[\rho_n P_j]P_j]\\ \nonumber
    & \sim \frac{2}{d\, N\, M\, L}\sum_{n,m,l}\frac{1}{\mathrm{Pr}(j_{n,m})}\mathrm{Tr}[\rho_n P_{j_{n,m}}]e^{j_{n,m}}_l.
\end{align}
$\mathrm{Pr}(j_{n,m})=\mathrm{Tr}[\rho_n P_{j_{n,m}}]^2/d$ is the relevance distribution telling us how relevant a certain Pauli operator is in determining the state fidelity for a particular $n$.

We can then use as a reward this fidelity estimator built from the result of local Pauli measurements on the final episode state with the initial state drawn uniformly at random in the set of states considered. This is different from direct fidelity estimation for gates and indeed the estimator is just a lower bound to the process fidelity. However, the hope is that this could be enough for learning since increasing the lower bound also increases the actual function we want to maximize and estimating such lower bound is more convenient in terms of resources since we sample from only $2d$ input states. For the single-shot protocol we just need to fix $N=M=L=1$ and the reward is $\frac{2}{d}\frac{1}{\mathrm{Pr}(j_{n,m})}\mathrm{Tr}[\rho_n P_{j_{n,m}}]e^{j_{n,m}}_l$.
\\
\\
\textbf{Note that,} for few qubits it is also possible to just select all the $2d$ initial states ending up with a $2d\times L$ shots protocol when $M=1$. Or even choose all input and all Pauli ending up with a $2d\times d^2\times L$ shots protocol.

\subsection{Gate engineering: 0-fidelity reward strategy}\label{app:0fid}
Finally, we formulate the  0-fidelity~\cite{PhysRevResearch.3.033031} estimator of the process fidelity as single-shot reward strategy for an RL agent.

The 0-fidelity between a unitary gate and a quantum channel is expressed as
\begin{equation}\label{0fid1}
    d\,F_0=1/d^2 \sum_{i,j=1}^{d^2}\mathrm{Tr}[\Lambda(\rho_i) P_j] \mathrm{Tr}[\Gamma(\rho_i)P_j],
\end{equation}
where $\Lambda$ is the desired channel and $\Gamma$ the actual one. Here we consider $\{P_k\}$ to be the Pauli operators basis.
We can rewrite this expression as
\begin{equation}\label{0fid2}
    F_0=\sum_{i,j=1}^{d^2}\mathrm{Pr}(i,j)\frac{\mathrm{Tr}[\Gamma(\rho_i)P_j]}{\mathrm{Tr}[\Lambda(\rho_i)P_j]}=\mathbb{E}_{n\sim Pr}[\frac{\mathrm{Tr}[\Gamma(\rho_{i_n})P_{j_n}]}{\mathrm{Tr}[\Lambda(\rho_{i_n})P_{j_n}]}],
\end{equation}
i.e., as expectation value over the probability $\mathrm{Pr}(i,j)=1/d^3 \mathrm{Tr}[\Lambda(\rho_i) P_j]^2$. This can be estimated via
\begin{equation}\label{0fid21}
    F_0\sim \frac{1}{N}\sum_{n=1}^N[\frac{\mathrm{Tr}[\Gamma(\rho_{i_n})P_{j_n}]}{\mathrm{Tr}[\Lambda(\rho_{i_n})P_{j_n}]}].
\end{equation}

Finally, in order to get a binary reward, we need to estimate the expectation value $\mathrm{Tr}[\Gamma(\rho_{i_n})P_{j_n}]$ according to the statistics of the final experimental state $\Gamma(\rho_i)$.

We thus get
\begin{align}\label{0fid3}
    F_0 &=\sum_{i,j=1}^{d^2}\mathrm{Pr}(i,j)\frac{1}{\mathrm{Tr}[\Lambda(\rho_i)P_j]} \sum_{m=1}^{d} e^{j}_{m} \mathrm{Tr}[\Gamma(\ket{v^{j}_{m}}\bra{v^{j}_{m}})P_j]\\ \nonumber
    &=\mathbb{E}_{n\sim Pr}\mathbb{E}_{m\sim \mathrm{Tr}[\ket{v^{j_n}_{m}}\bra{v^{j_n}_{m}}}[\frac{e^{j_n}_m}{\mathrm{Tr}[\Lambda(\rho_{i_n})P_{j_n}]}],
\end{align}
which can be estimated via
\begin{equation}\label{0fid31}
    F_0\sim \frac{1}{N\,M}\sum_{n=1}^N\sum_{m=1}^M[\frac{e^{j_n}_m}{\mathrm{Tr}[\Lambda(\rho_{i_n})P_{j_n}]}].
\end{equation}
From this last expression we see that the reward $\frac{e^{j_n}_m}{\mathrm{Tr}[\Lambda(\rho_{i_n})P_{j_n}]}$ can be used in the RL algorithm.

The complete protocol for the 0-fidelity estimation is then:
\begin{itemize}
    \item select $N$ pairs of indices $(i_n,j_n)$ according to the relevance distribution, this sets the measurement Pauli (measurement settings) and the input state
    \item measure the measurement Pauli for each of the input states
    \item for single-shot reward in the RL algorithm it is enough to just sample $N=M=1$ at each iteration.
\end{itemize}

\section{Black-box optimization algorithms}\label{app:blackbox}

In the main text, we compare the results obtained via model-free RL with other well-known black-box optimization methods. In this appendix, we summarize the basics of the methods considered.

\subsection{Bayesian Optimization}\label{subsec:BO}
Bayesian Optimization (BO) is another black-box optimization method which is suitable in the presence of noisy evaluations for the function whose extrema we are interested in obtaining~\cite{shahriari2015taking}.

Consider $f(x)$ as the function to minimize, $y(x)$ its noisy realization, and $x$ the vector of parameters over which to optimize. At iteration $t$, given all previous observations $\{ x_i,y_i\}_{i=1,...,t}$, we update a surrogate, probabilistic model for the objective function.
This is usually a Gaussian process whose mean approximates the optimization landscape, i.e., $f(x)$. The optimization is done via a cheap acquisition function based on the posterior distribution for sampling the next point.

When using single-shot experiments, BO has been employed for quantum control by properly modeling the binomial measurement noise~\cite{PRXQuantum.1.020322}. However appealing, this strategy seems to reach a bottleneck when the function to estimate involves several probabilities for which a different surrogate model is required. Here we thus employ standard BO with one random-forest-family (Extra Trees) surrogate. While this cannot learn with single-shot measurement -- in contrast to~\cite{PRXQuantum.1.020322} -- it can learn with few shots. Thus, when comparing to the single-shot RL agents, the objective function is obtained by averaging single-shot results over 500 episodes, i.e., using a 500-shot fidelity estimate.

\subsection{Evolution strategies}\label{subsec:evolutionary}

Evolutionary algorithms are a wide class of optimization methods in which best solutions are picked from a candidate solutions group and iteratively improved via small changes over many generations. These techniques have found widespread use in quantum control and error mitigation~\cite{PhysRevLett.124.220501,PhysRevA.90.032310,9504796,PhysRevA.88.052306,zheng2022optimal,pack2026benchmarking,dq3h-1k35}.

Here we employ the covariance matrix adaptation evolution strategy (CMA-ES)~\cite{hansen2003reducing}.
In this method, new candidate solutions are sampled from a multivariate normal distribution whose parameters are then updated iteratively until convergence. CMA-ES has been shown to perform well with respect to other black-box optimization methods and is considered well performing on non-convex, noisy objective functions~\cite{hansen2010comparing}.

We use the python package \textit{cma}~\cite{Hansen2019} and the recently introduced Learning Rate Adaptation \mbox{(LRA-)CMA}~\cite{nomura2025cma,10.1145/3583131.3590358} -- present in the package -- that has been shown to handle noise effectively.

As for BO, also CMA-ES does not work with single-shot fidelity. Thus, when comparing to the single-shot RL agents, the objective function is once again obtained by averaging single-shot results over 500 episodes, i.e., using a 500-shot fidelity estimate.

\section{Extra results}
We collect here additional results that were omitted in the main text.
\subsection{Hada4 and Hada10 tasks}\label{app:Hada4}
\begin{figure}[H]
\centering
\begin{subfigure}[t]{0.49\linewidth}
    \centering
    \includegraphics[width=\linewidth]{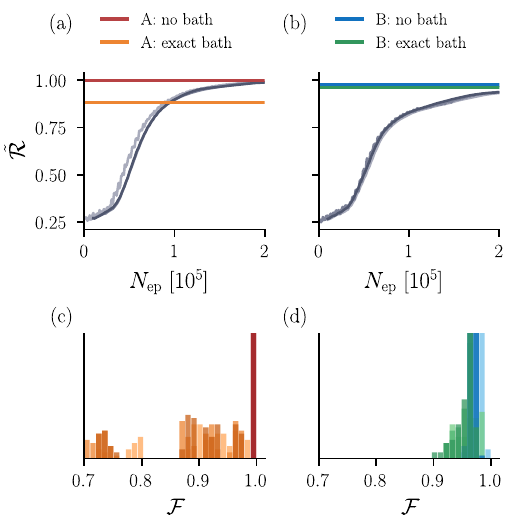}
    \caption{\label{fig:learning_Hada4_v2} Exact fidelity agents.}
\end{subfigure}
\hfill
\begin{subfigure}[t]{0.49\linewidth}
    \centering
    \includegraphics[width=\linewidth]{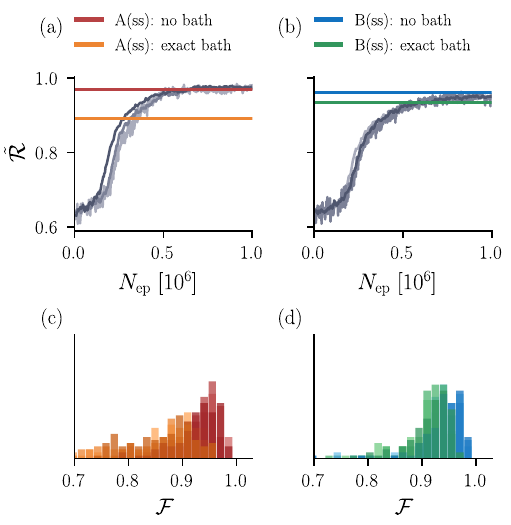}
    \caption{\label{fig:learning_Hada4ss_v2} Single-shot agents.}
\end{subfigure}
\caption{\label{fig:learning_Hada4}
Similar to Figs.~\ref{fig:Hada10}(a)-(d) with target Hada4: (a) exact fidelity agents and (b) single-shot agents.}
\end{figure}

\begin{figure}[H]
\centering
\begin{subfigure}[t]{0.49\linewidth}
    \centering
    \includegraphics[width=\linewidth]{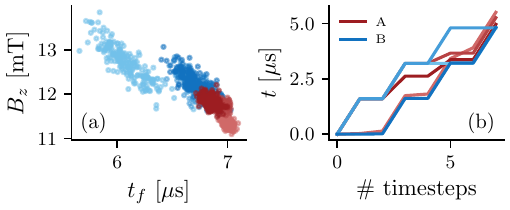}
    \caption{\label{fig:scatter_Hada4_v2} Exact fidelity agents.}
\end{subfigure}
\hfill
\begin{subfigure}[t]{0.49\linewidth}
    \centering
    \includegraphics[width=\linewidth]{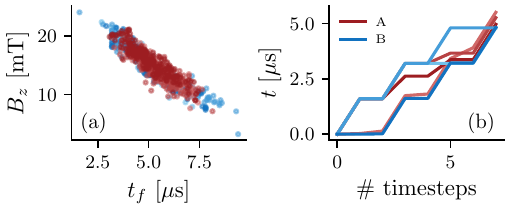}
    \caption{\label{fig:scatter_Hada4ss_v2} Single-shot agents.}
\end{subfigure}
\caption{\label{fig:scatter_Hada4}
Similar to Figs.~\ref{fig:Hada10}(g) and (h) with target Hada4: (a) exact fidelity agents and (b) single-shot agents.}
\end{figure}

\begin{figure}[H]
\centering
\includegraphics[width=0.5\linewidth]{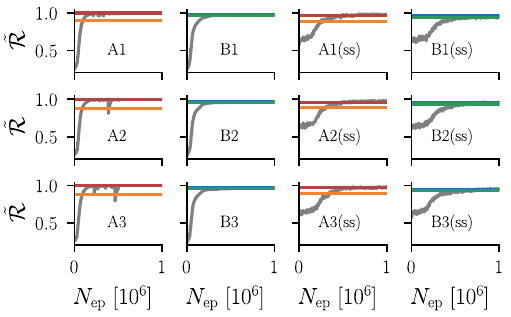}
\caption{\label{fig:all_learning_Hada4}
Learning curves for the different agents trained on the Hada4 target. Horizontal lines indicate the fidelity obtained after testing either without bath (red/blue) or with the exact bath (orange/green). }
\end{figure}

Figs.~\ref{fig:learning_Hada4} and \ref{fig:scatter_Hada4} parallel what we have shown in section~\ref{sec:structure-bath} of the main text for the Hada4 task. In particular, we see that agents B learn to take the structured spin-bath into account and choose shorter pulses and higher magnetic fields with respect to agents A trained in the absence of the spin-bath.
Fig.~\ref{fig:all_learning_Hada4} shows the learning curves and evaluation results for all the trained agents for the Hada4 task.

\begin{figure}[H]
\centering
\includegraphics[width=0.5\linewidth]{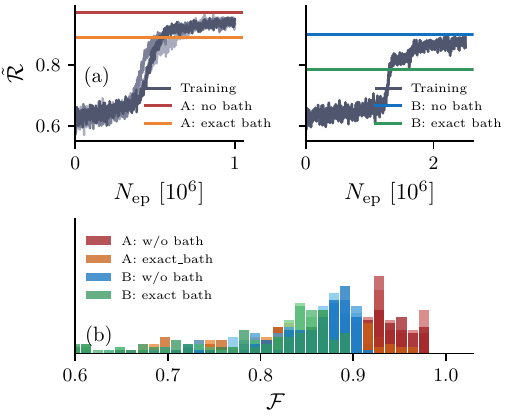}
\caption{\label{fig:learning_Hada10ss_v2} Similar to Figs.~\ref{fig:Hada10}(a)-(d) with single-shot agents for Hada10 task.}
\end{figure}
\begin{figure}[H]
\centering
\includegraphics[width=0.5\linewidth]{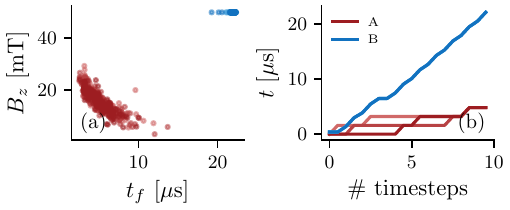}
\caption{\label{fig:scatter_Hada10ss_v2} Similar to Figs.~\ref{fig:Hada10}(g) and (h) with single-shot agents for Hada10 task.}
\end{figure}

Fig.~\ref{fig:learning_Hada10ss_v2} and Fig.~\ref{fig:scatter_Hada10ss_v2} parallel the analysis done for the Hada10 task with exact fidelity reward in the main text but use a single-shot reward. We see that with the single-shot reward the quality of the results is far from what we obtained with the exact fidelity and agents B(ss) are not able to find shorter pulse sequences (while still preferring higher magnetic field magnitudes).

\begin{figure}[H]
\centering
\includegraphics[width=\linewidth]{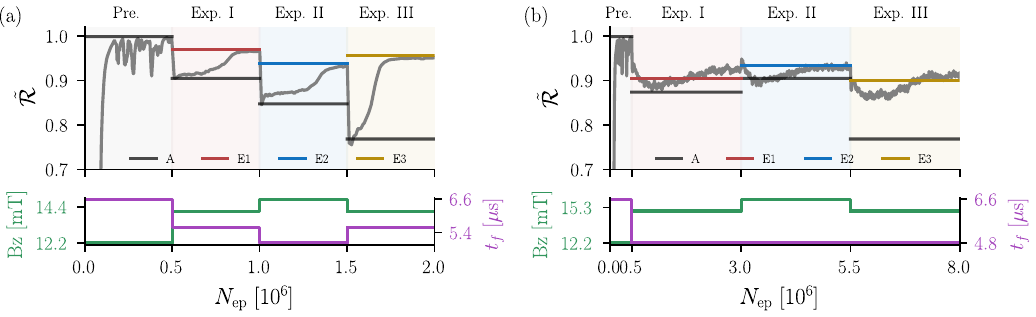}
\caption{\label{fig:pre-training_Hada10}
Similar to Fig.~\ref{fig:pre-training_Hada4_combined} for Hada10
 }
\end{figure}

Finally, Fig.~\ref{fig:pre-training_Hada10} shows the result of pre-training in the absence of a spin-bath and then continuing learning in the presence of different spin-baths for the Hada10 task in parallel with the results for the Hada4 task shown in the main text. Here we see that, also with single-shot reward agents B(ss) (right panel) are once again able to learn shorter pulse sequences when aided by a pre-training phase.

\section{Benchmarking: additional results}\label{app:blackbox_results}
\begin{figure}[H]
\centering
\begin{subfigure}[t]{0.49\linewidth}
    \centering
    \includegraphics[width=\linewidth]{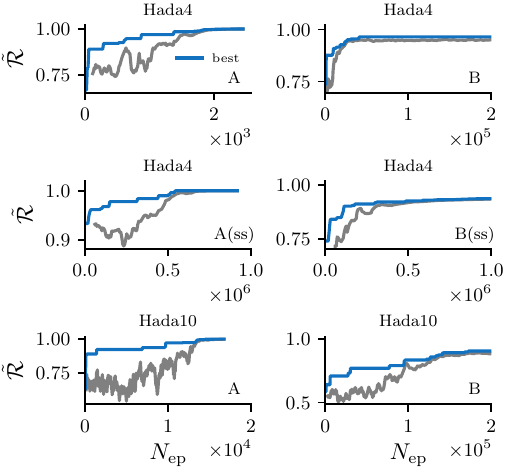}
    \caption{\label{fig:CMA} CMA-ES optimization.}
\end{subfigure}
\hfill
\begin{subfigure}[t]{0.49\linewidth}
    \centering
    \includegraphics[width=\linewidth]{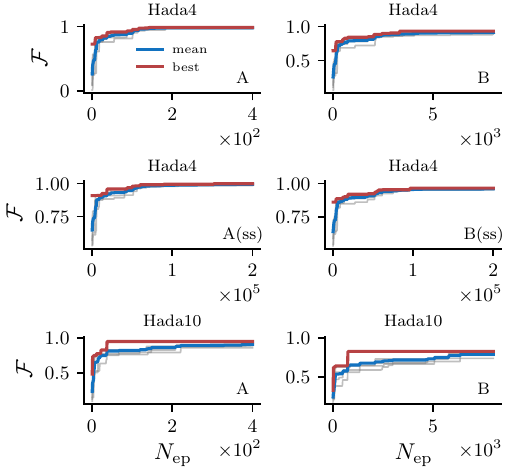}
    \caption{\label{fig:BO} BO optimization.}
\end{subfigure}
\caption{\label{fig:optimization}
CMA-ES (left) and BO (right) optimization for the agents from Tab.~\ref{tab:benchmarks}. For CMA-ES a smoothing with window 10 was used.}
\end{figure}

We report here the convergence curves for the black-box optimization methods used as benchmarks for our agents.
Fig.~\ref{fig:CMA} and Fig.~\ref{fig:BO} show the results of the CMA and BO algorithms respectively.

\subsection{Bloch sphere trajectories analysis}\label{app:BlochSphere}
\begin{figure}[H]
\centering
\includegraphics[width=0.75\linewidth]{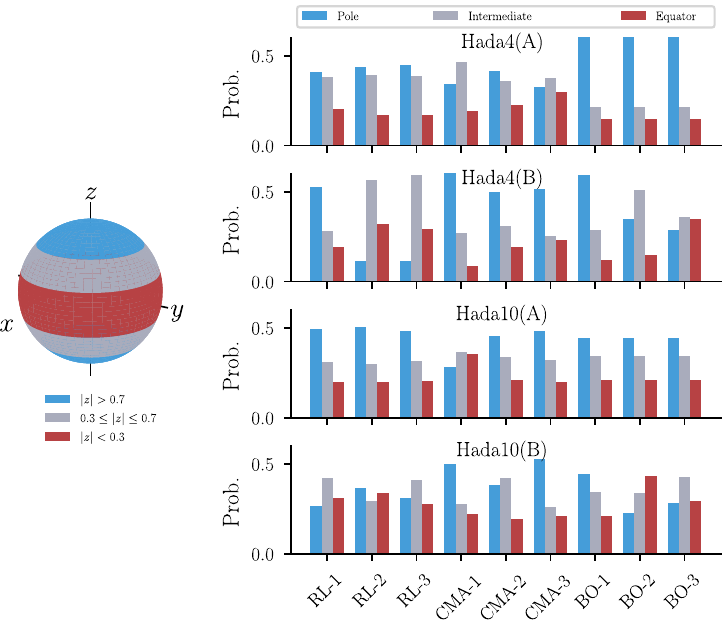}
\caption{\label{fig:bloch_histo}
Classification of the learned trajectories on the Bloch sphere for the Hada4 and Hada10 targets. The histograms show how often the trajectories generated by RL, CMA-ES, and BO remain near the poles, pass through intermediate regions, or approach the equator, based on the $z$-projection of the nuclear spin. We define equatorial regions by $|\langle \sigma_z \rangle| < z_{\mathrm{eq}}$ and polar regions by $|\langle \sigma_z \rangle| > z_{\mathrm{pole}}$, with $z_{\mathrm{eq}} = 0.3$ and $z_{\mathrm{pole}} = 0.7$.}
\end{figure}

We furthermore report here an analysis of the time spent by the agents' policies, for the synthesis of the Hadamard gate, in different parts of the Bloch sphere. This is compared with the protocols found via CMA-ES and BO. Fig.~\ref{fig:bloch_histo} shows how often the trajectories from different optimization methods remain near the poles, pass through intermediate regions, or approach the equator -- with the different regions defined as in the figure caption -- based on the $z$-projection of the nuclear spin.

\section{Hyperparameters}\label{app:hyperpa}

We report the hyperparameters used for RL agents in Tab.~\ref{tab:ppo_hyperparams} and black-box benchmarks in Tabs.~\ref{tab:cmaes_hyperparams} and \ref{tab:bo_hyperparams}. Except where noted, defaults of the respective packages were used. No systematic hyperparameter optimization was performed beyond manual tuning, so the reported performance should not be considered optimal and could likely be improved with a more extensive search. For a fair comparison, all three optimizers share the same objective (infidelity $1-\mathcal{F}$) and evaluation budget. For the black-box methods the fidelity was obtained by sampling over 20 baths for agent B and over 500 baths and shots for agents A(ss) and B(ss).

\paragraph{Reinforcement Learning Hyperparameters}
Training used PPO as implemented in \texttt{stable-baselines3}, with a Gaussian policy. Values not listed correspond to library defaults.
\begin{table}[H]
\footnotesize
\centering
\begin{tabular}{ll}
\hline
\textbf{Hyperparameter} & \textbf{Value} \\
\hline
Algorithm & PPO \\
Number of steps ($n_{\text{steps}}$) & 2000 \\
Batch size & 200 \\
Number of epochs ($n_{\text{epochs}}$) & 10 \\
Clipping range ($\epsilon$) & 0.2 \\
Learning rate & $5 \times 10^{-4}$ \\
Entropy coefficient & 0.005 \\
Discount factor ($\gamma$) & 0.9996 \\
\hline
\end{tabular}
\caption{PPO hyperparameters.}
\label{tab:ppo_hyperparams}
\end{table}

\paragraph{CMA-ES Hyperparameters}
The optimization was performed using the \texttt{cma} package.
\begin{table}[H]
\footnotesize
\centering
\begin{tabular}{ll}
\hline
\textbf{Hyperparameter} & \textbf{Value} \\
\hline
Initial mean & $\mathcal{U}(-1, 1)^N$ (seed 0) \\
Initial step size & $1$ \\
Function tolerance & $5\times 10^{-3}$ \\
Evaluation budget & same as RL agents \\
Search-space dimension & $N = n_\mathrm{actions} \times n_\mathrm{steps}$ \\
Parameter bounds & $[-1, 1]^N$ \\
\hline
\end{tabular}
\caption{CMA-ES hyperparameters.}
\label{tab:cmaes_hyperparams}
\end{table}

\paragraph{Bayesian Optimization Hyperparameters}
The optimization was performed using \texttt{scikit-optimize}.
\begin{table}[H]
\footnotesize
\centering
\begin{tabular}{ll}
\hline
\textbf{Hyperparameter} & \textbf{Value} \\
\hline
Surrogate model & Extra Trees \\
Acquisition function & Expected Improvement \\
Acquisition optimizer & sampling \\
Initial points & 100 \\
Iterations per run & 400 \\
Batch size & 1 \\
Independent runs & 5 (seeds 0--4) \\
Search-space dimension & $N = n_\mathrm{actions} \times n_\mathrm{steps}$ \\
Parameter bounds & $[-1, 1]^N$ \\
\hline
\end{tabular}
\caption{Bayesian optimization hyperparameters.}
\label{tab:bo_hyperparams}
\end{table}

\end{document}